# Benchmarking Over The Grid Using Atlas Virtual Organization.

**May 2007**

# Technical Report

**Ioannis Kouvakis[1] and Fotis Georgatos[2]**

[1] **Student of Univerity of Aegean, Department of Mathematics,**
gkouvakis@hep.ntua.gr
*(special Thanks to my thesis advisor Dr. Ioanis Gialas for giving me the opportunity to do research on Grid)*

[2] **University of Cyprus, Department of Informatics,**
fotis@mail.cern.ch

# Introduction

Grids include heterogeneous resources, which are based on different hardware and software architectures or components. In correspondence with this diversity of the infrastructure, the execution time of any single job, as well as the total grid performance can both be affected substantially, which can be demonstrated by measurements.

# Approach

In order to see this heterogeneity of the grid we send a simple script job over some sites of the Grid using the Atlas V.O. The script had 2 parts. The first part was taking information from the worker node that the script was running, such as Kernel Version, Linux Distribution, Environment Variables, Packages and Kernel Messages and also Hardware Information concerning Cpu Model / Vendor / Speed, Memory Size, Hard Disks and other Media Details. The second part of the script was downloading / compiling and running the Lmbench Benchmarking Suite that it was analyzing the performance of the current node.

The job was send several times over different worker nodes of each site. Based on the fact that we have homogeneity inside each site, which it was true in almost all the sites, we were able to get some statistical information and compare those sites.

# Results

The results of this approach benefits all grid users, since optimization of the system as a whole, can lead to direct and indirect advantages for everyone, in terms of both individual AND total job throughput. What we want specifically to demonstrate is, that bu neglecting benchmarking and resource characterization, enormous amount of grid resources can be waisted or sub-optimally exploited. For example, the systems that are best on floating point of a given algorithm, say 64bit operations, are not always the ones that are optimal on memory transfers, and vice versa.

All the results from the script and the lmbench can be downloaded from [http://www.kouvakis.gr/lang_en/projects/grid.html](http://www.kouvakis.gr/lang_en/projects/grid.html). Also you can download the script and excel worksheets that were used to get all the pies and the statistical tables.

The information we got from the first part of the script, helped us make the following pies concerning cpu models that a job can run into and the numbers of cpu each worker node has. That information is per Job (projected) meaning that we had to resort to a projection technique, since benchmarking all of EGEE is impossible. Projection involves taking a number of measurements of a site, which is dependent on its size, so that we can obtain enough information to be able to deduce conclusions about its characteristics. The technique has not yet been verified in a scientific manner, but we do consider the graphs representative of reality or, near enough. Additional the graphs per Site is the percentage of the total sites we measure and not all as we have difficulties on taking results from sites that have been down that period or having other problems consering bugs on the middleware.

In the first pies we can see the percentage per job (pojected) and per site that use Intel or AMD processors and we conclude that the grid is based moslty on Intel as the 75% of it is Intel models.

**CpuVendor per Job (projected*)**
- AMD 25%
- Intel 75%

**CpuVendor per Site**
- AMD 23%
- Intel 77%

Also we can see the cpu model per job and per site, which is the number of jobslots or sites using the current model.

**CPU Model per Job (projected*)**

(Bar chart: # Jobs vs Cpu Model, ranging from 0 to 7000, with various AMD Athlon, AMD Opteron, Dual Core AMD Opteron, Intel Pentium, Intel Xeon, Intel XEON, Pentium III (Coppermine), and Itanium 2 models listed along the x-axis.)

[Chart: CPU Model per Site — bar chart showing # Sites vs Cpu Model]

In the Cpu Speed per cpus (in GHz) pie we can see the percentage of the speed that each job can theoretically use.

[Chart: Cpu Clock Speed per Job (projected*) — bar chart showing # Jobs vs Cpu Clock Speed (GHz)]

[Chart: Cpu Clock Speed per Site — bar chart showing # Sites vs Cpu Clock Speed (GHz)]

In the following charts we can see how much system memory can consume and also the total size of the swap memory. We have a major percentage of 4096MB and 2048. The swap memory is very confusing if we consider that not all the nodes are following the 2:1 Swap/Ram rule.

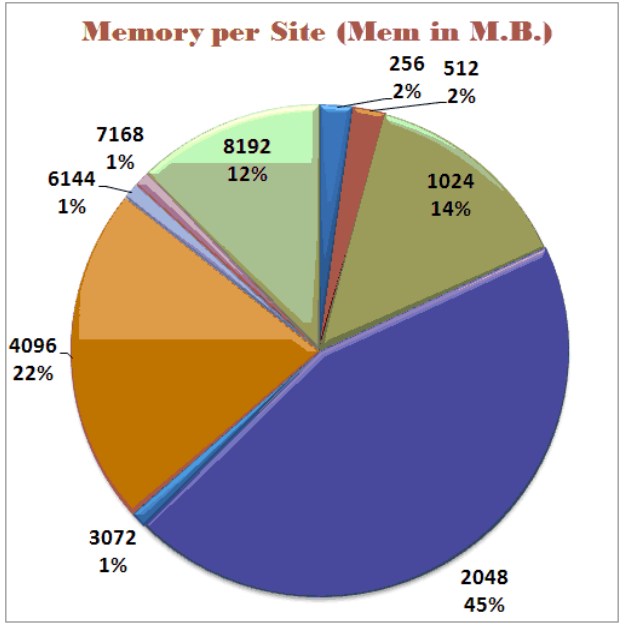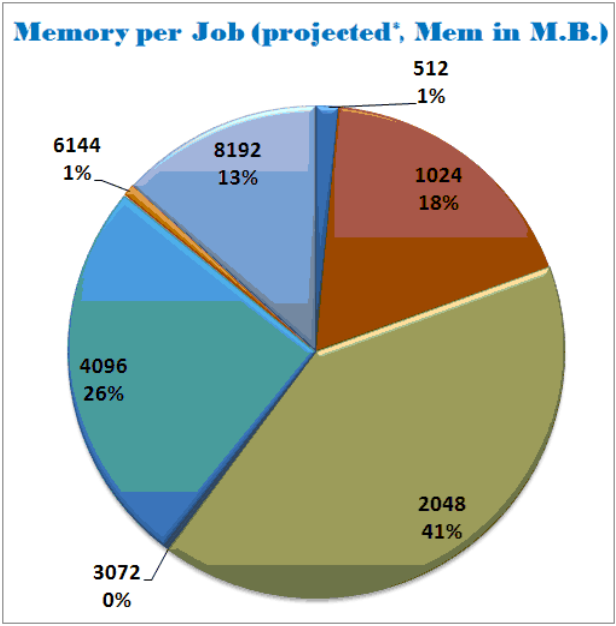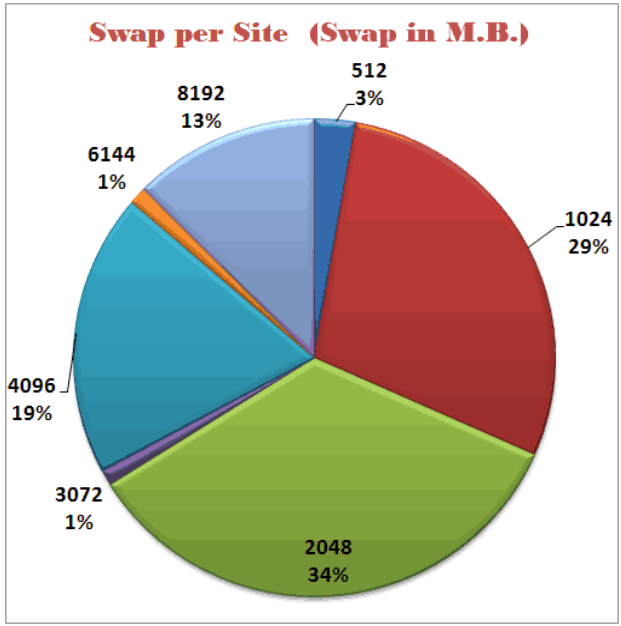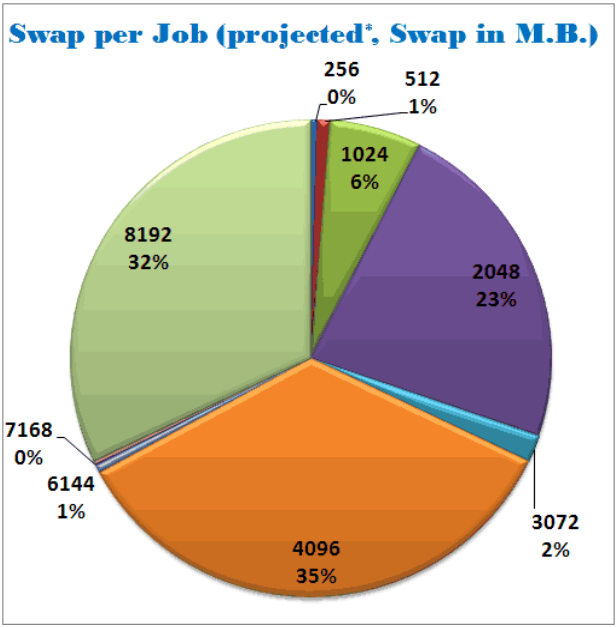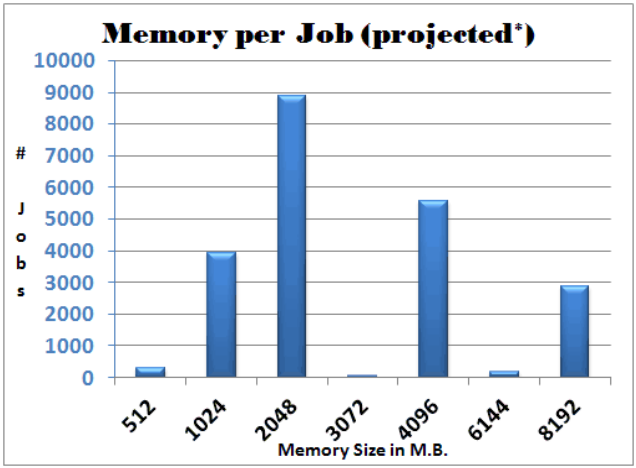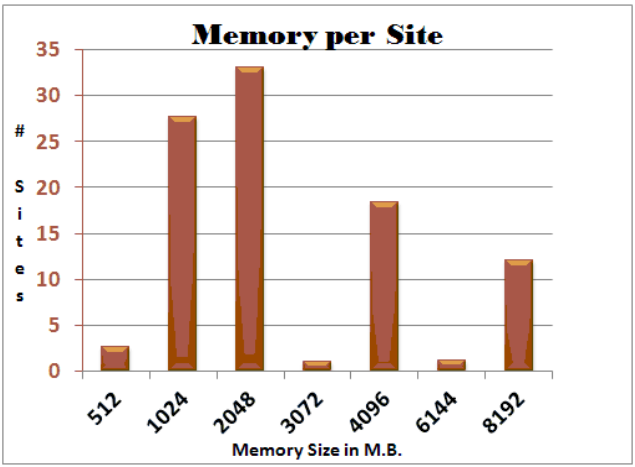

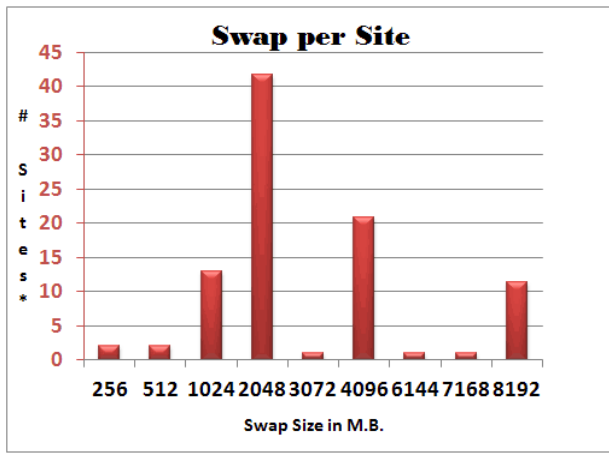
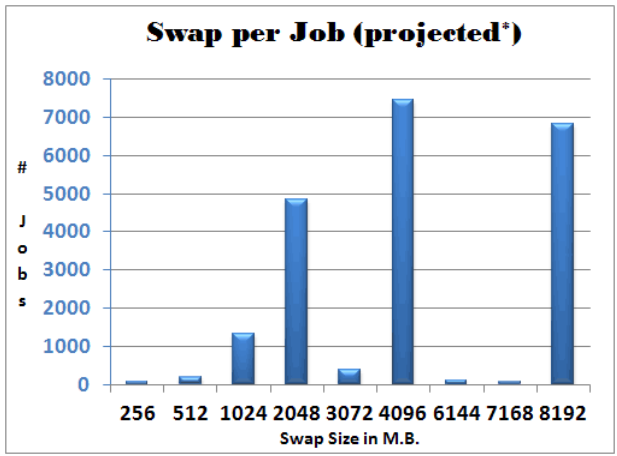
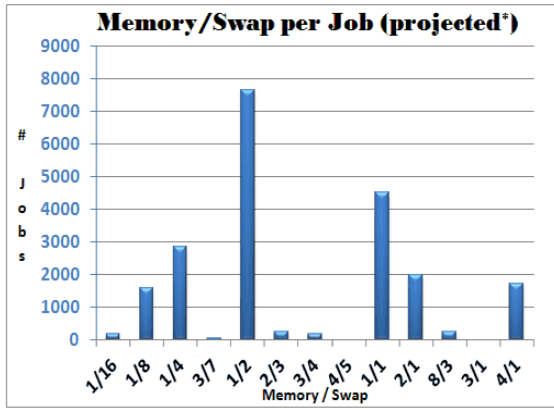
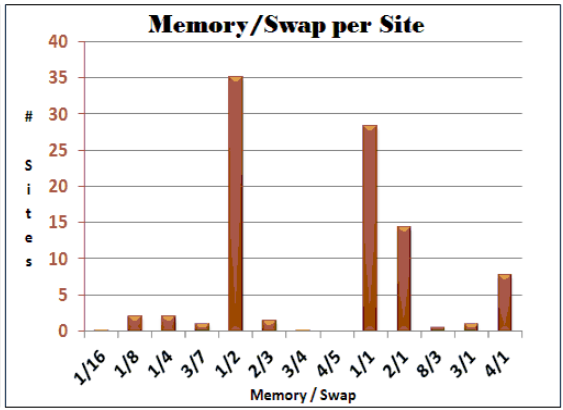

The following pies are showing the Linux Distribution and kernel version per site and per Job (projected). We can see a variety of kernel versions and that not all the sites have been updated to the latest. But we can see that a great percentage use take advantage of the smp technology of the cpu using the smp kernel version.

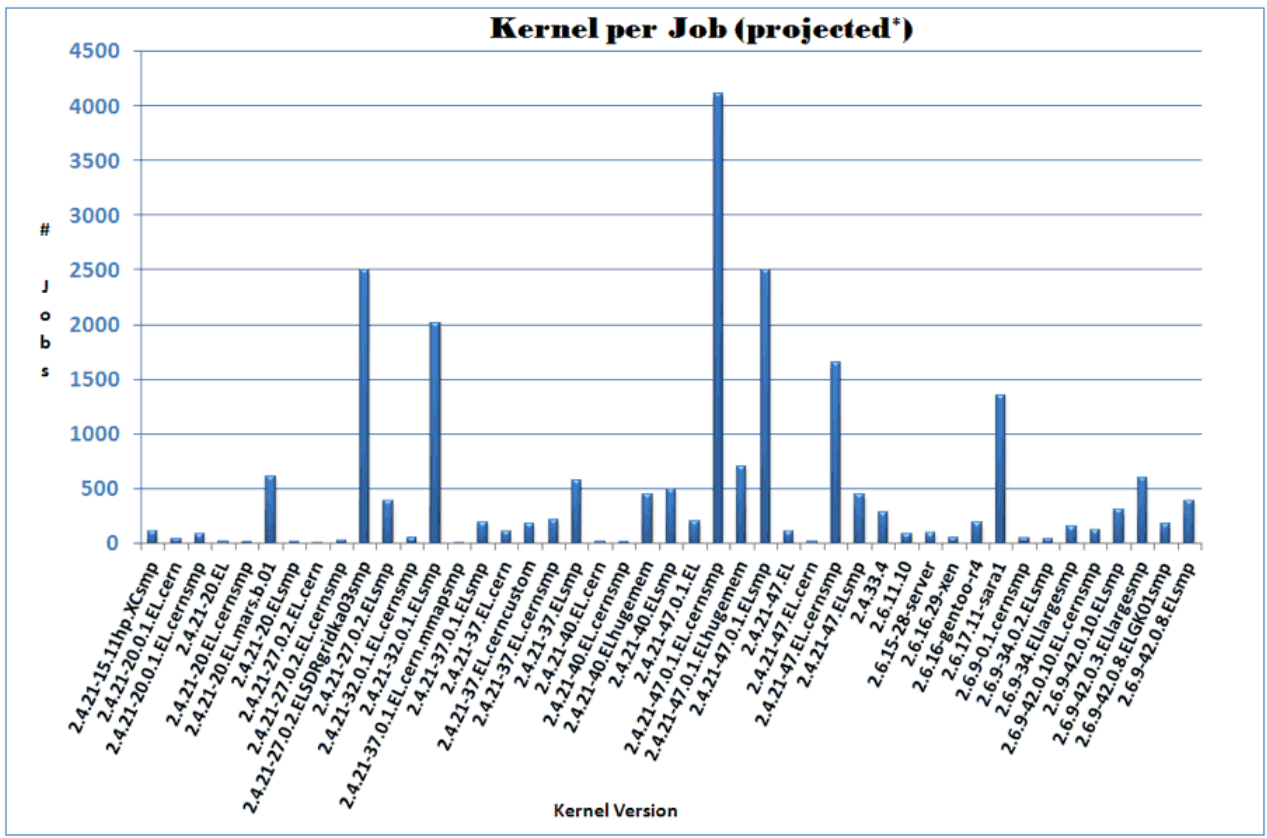

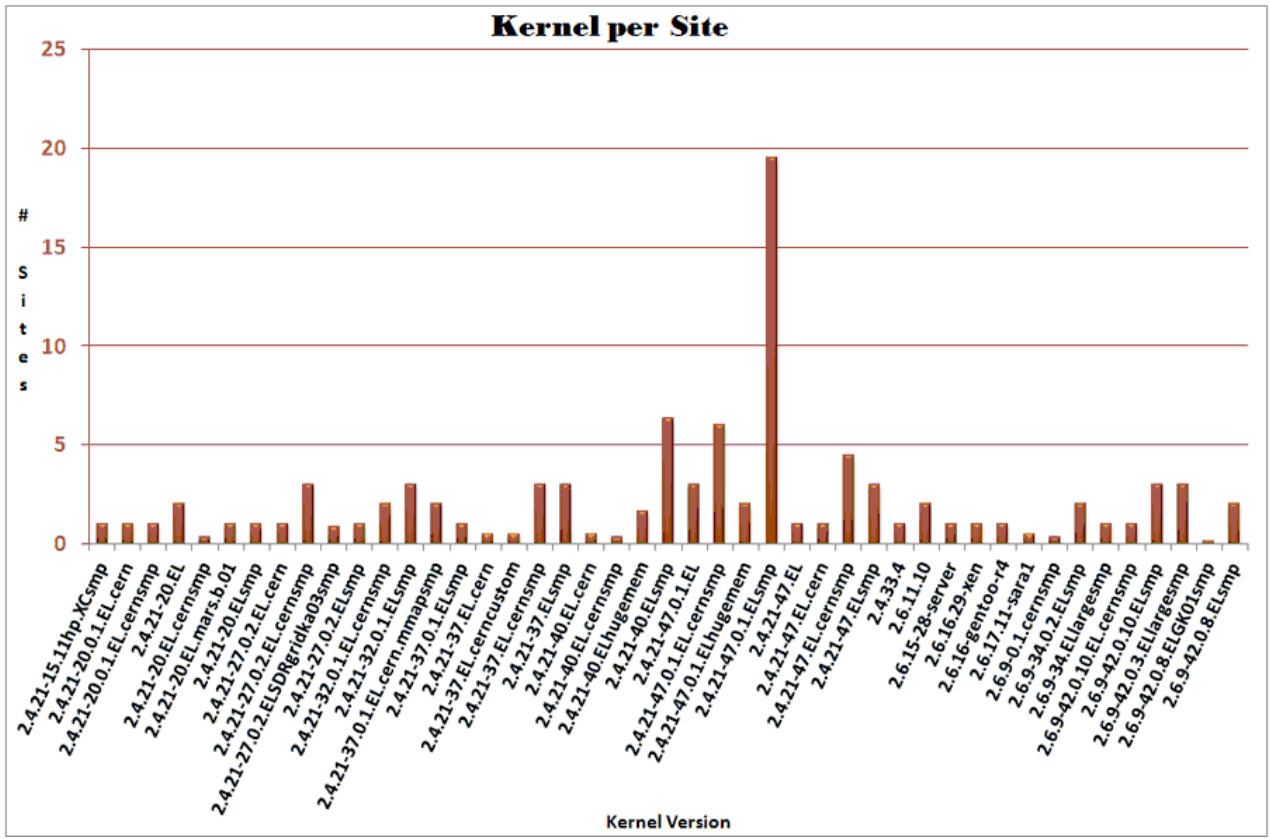

Only the 1/5 of the total sites are using the latest base for the kernel as the distribution that is the default for the grid did not supported kernels 2.6.

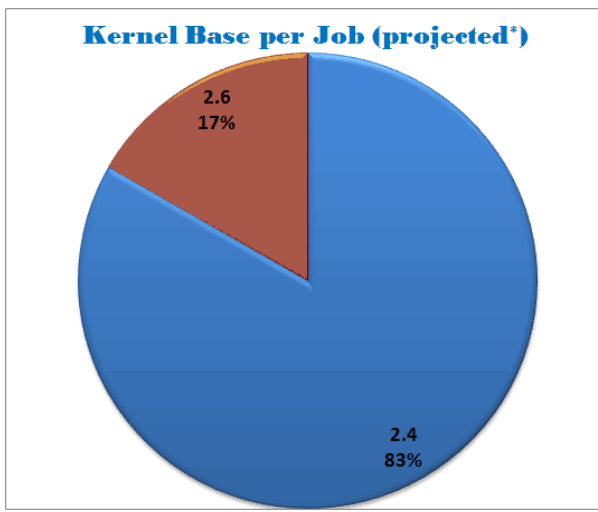
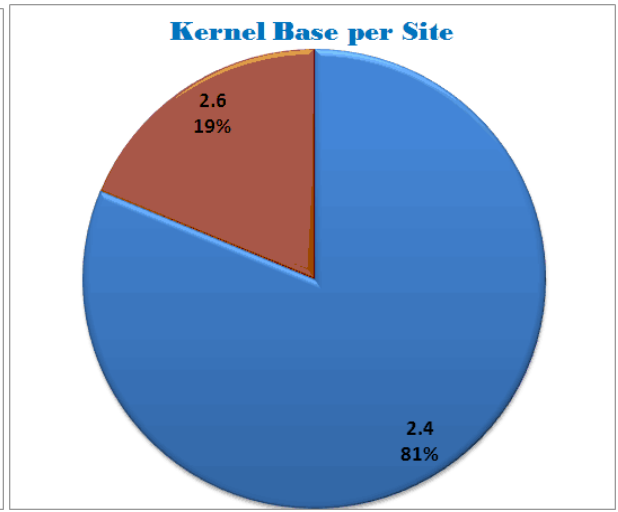

The distribution charts shows the intresting information that there are also other distribution than the Scientific Linux that are currently in the Grid.

The Middleware per Job (projected) and per Site charts were made by combining information from the script and the GSTAT site. A very small percentage is still using the old middleware and a great percentage use the Glite-3.0.0.

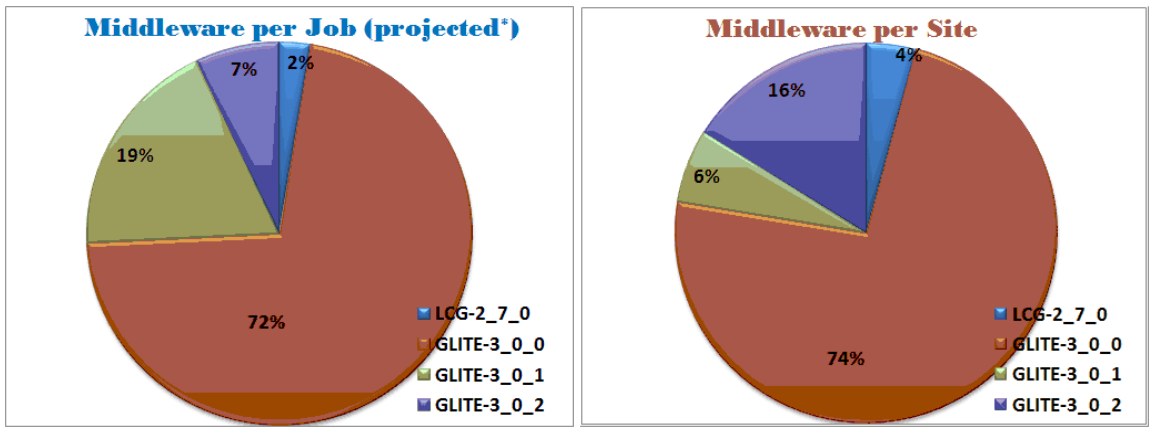

In the next chart we can see how many jobslots there are per site. This information is combined from GSTAT and lcginfosites and is not very accurate. There is no way to find out how many nodes does a site have and there is also the problem that a number of sites have internal heterogeneity, wich makes it difficult to find out how many jobslots they really have.

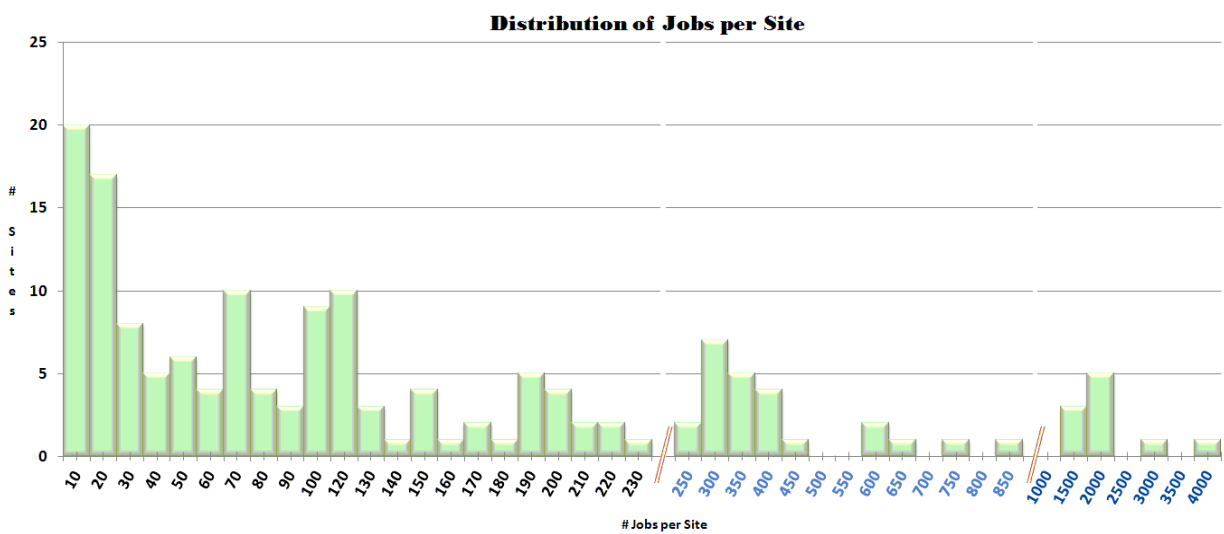

From the second part of the script we have the results of the lmbench, which it was compiled in every worker node to take advantage of the full capabilities of it. The results were exported on a csv format and imported on an excel worksheet so that it can be more analyzed and compared among the different sites.

One of the most critical benchmarks is the numerical operations measurements. Most researches are based on number results and operations between those numbers. The time needed for those operations has a major role on the grid. The lmbench tests four types of numbers such ass integer, float, double, and 64bit integers and many kind of operations such as add, multiply, division and for integers an additional, the mod.

In the following table we can see the average of float and double number operations which are the result of running the lmbench on different nodes on each site. The times are measured in nanoseconds and the smaller times are the better.

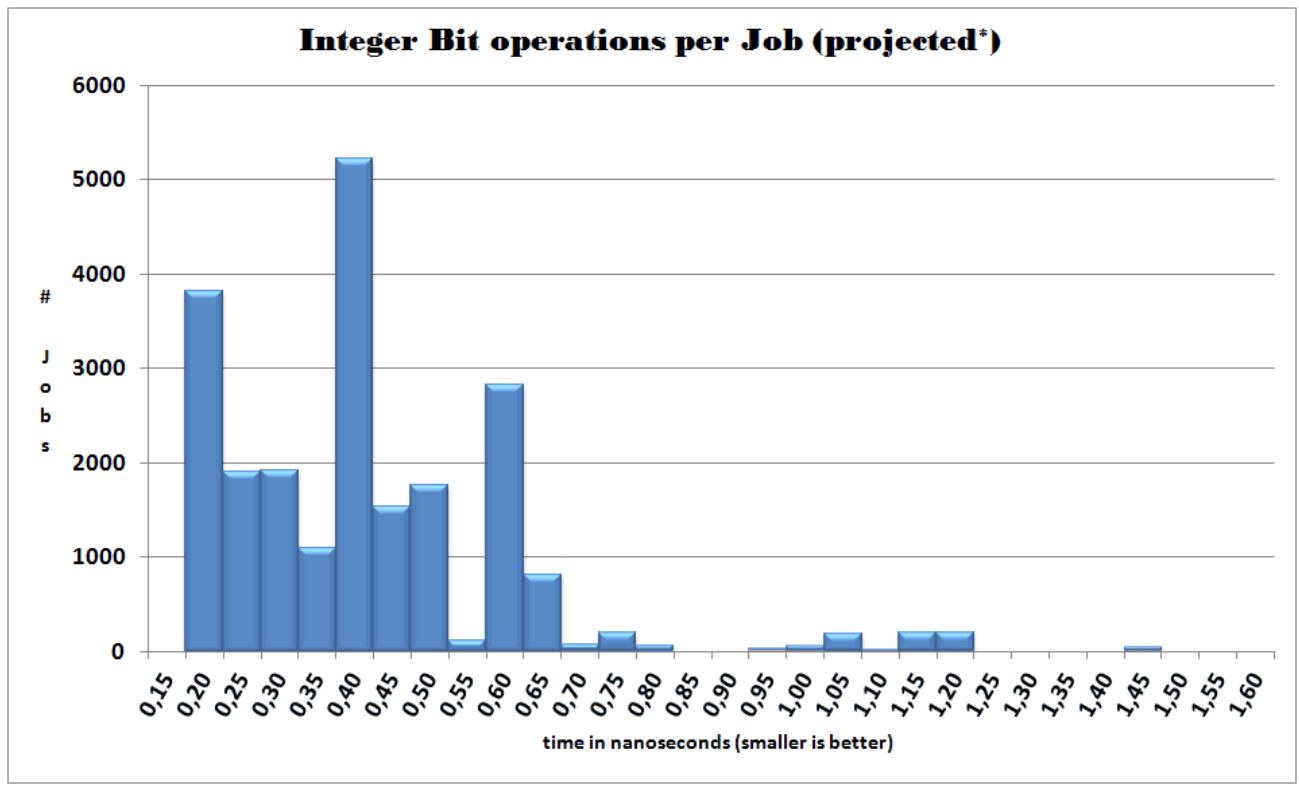

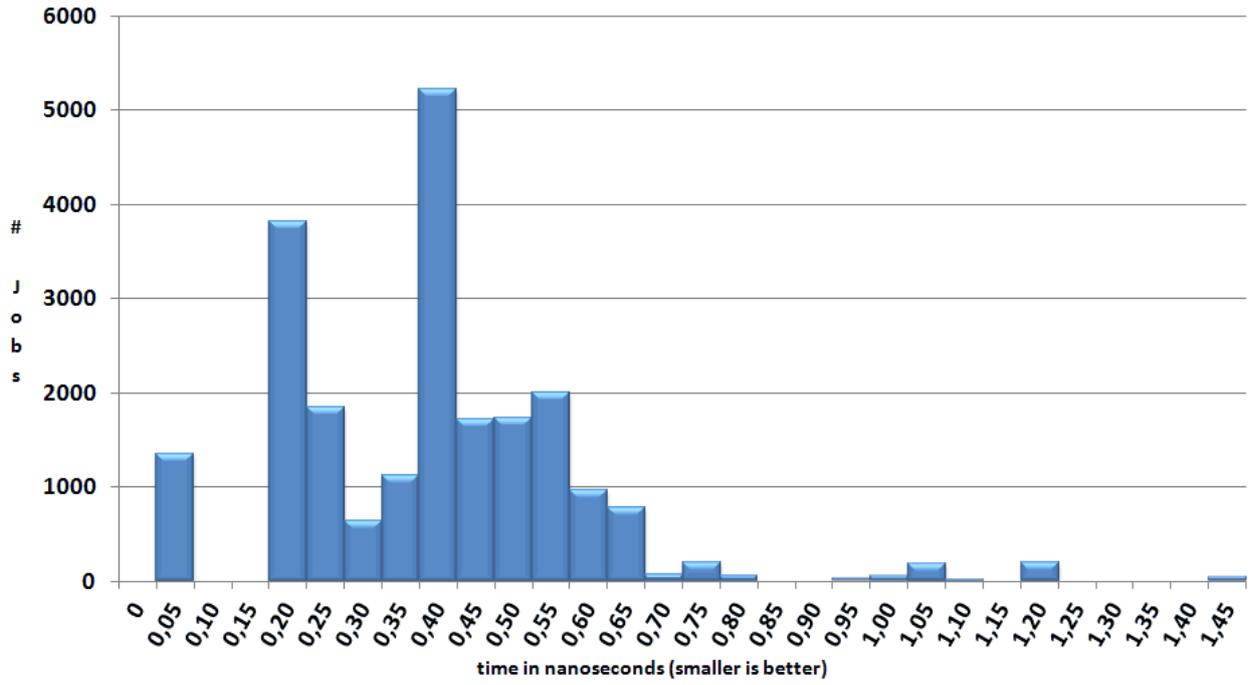
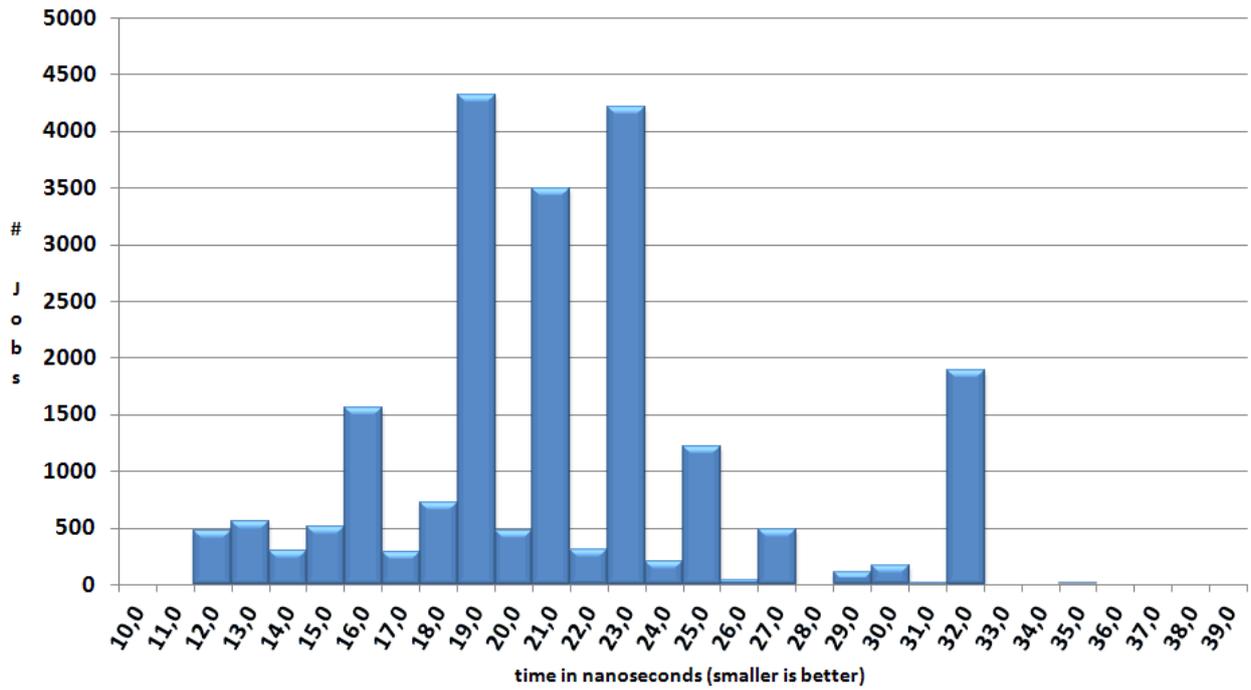

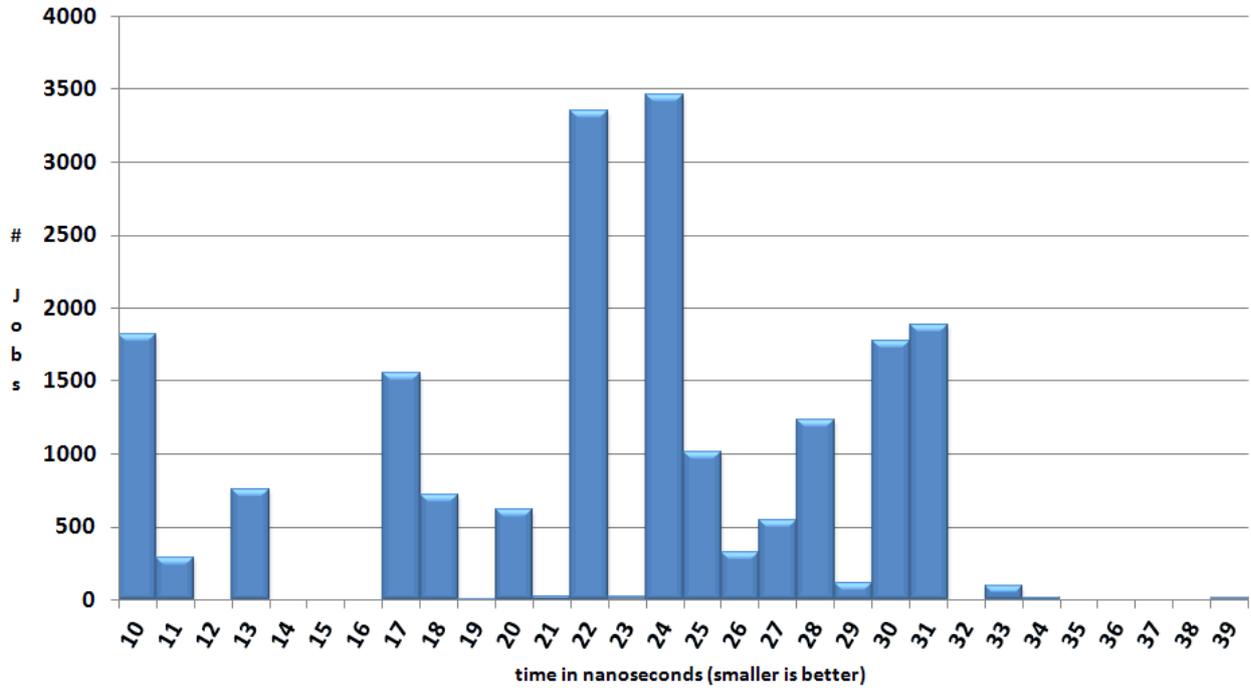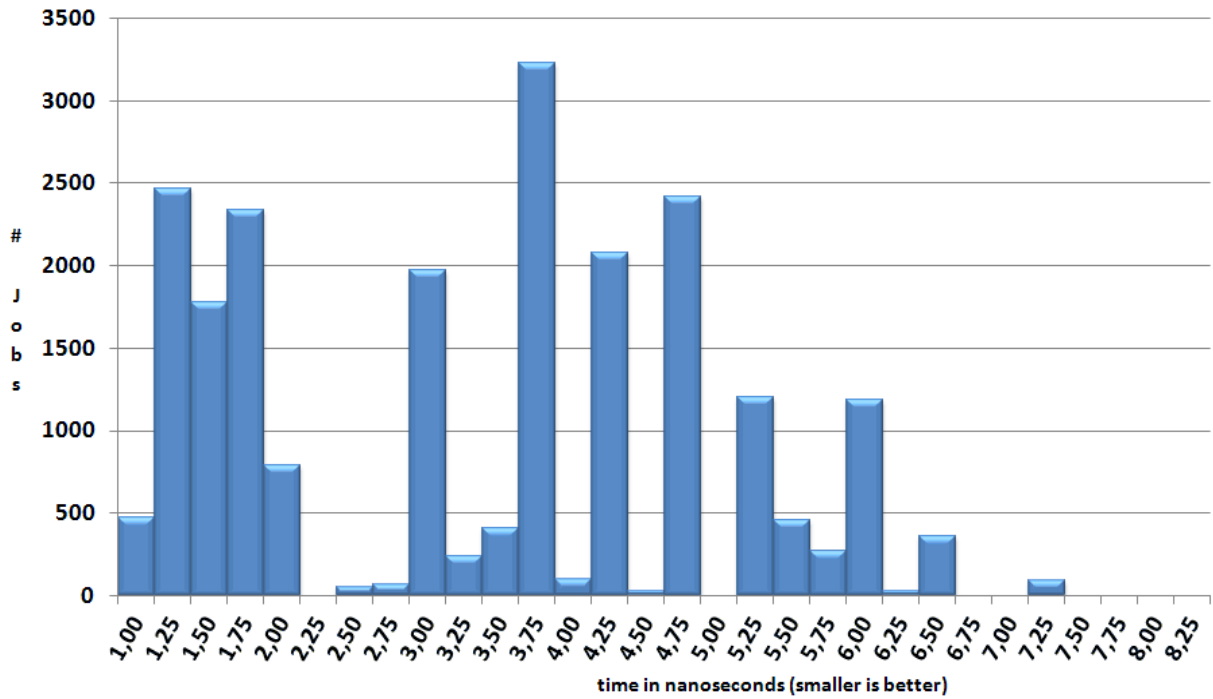

The same tables with averages for integer and 64bit integer numbers are shown below.

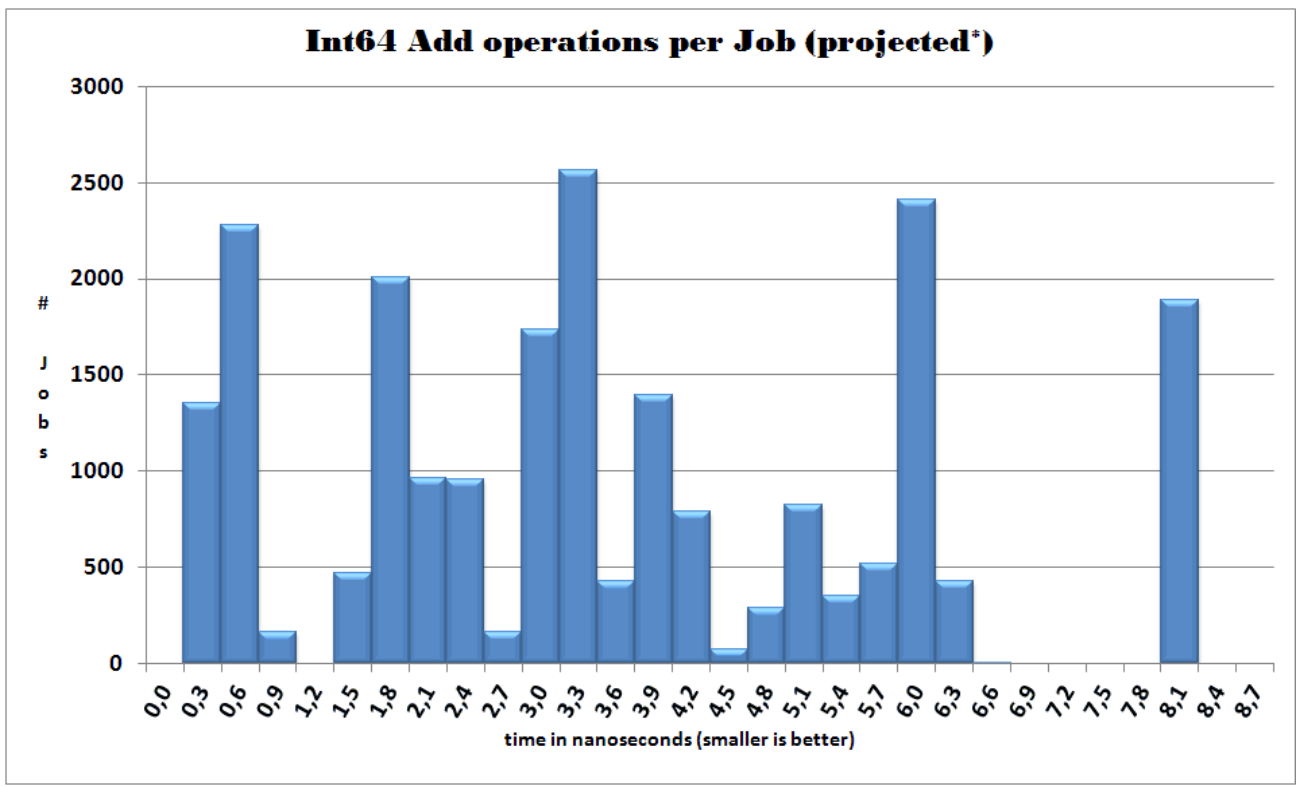

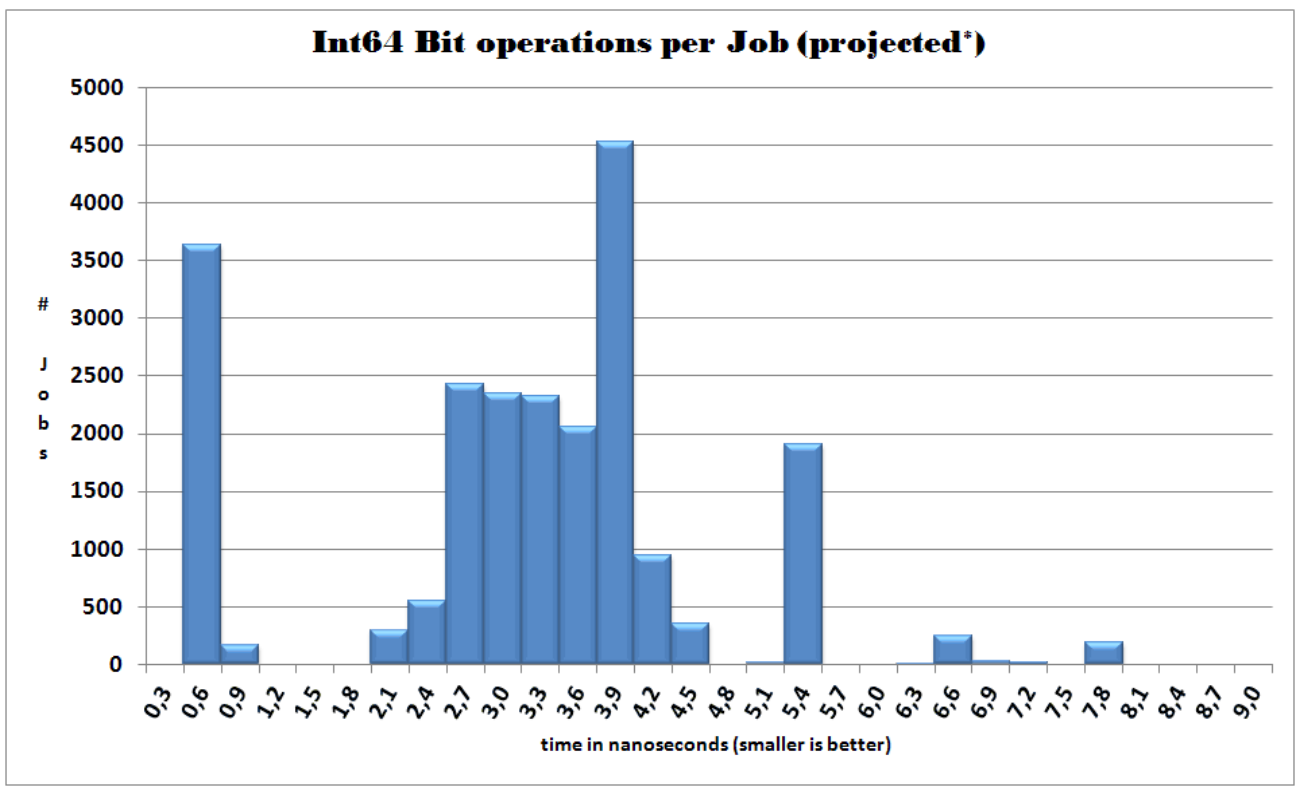

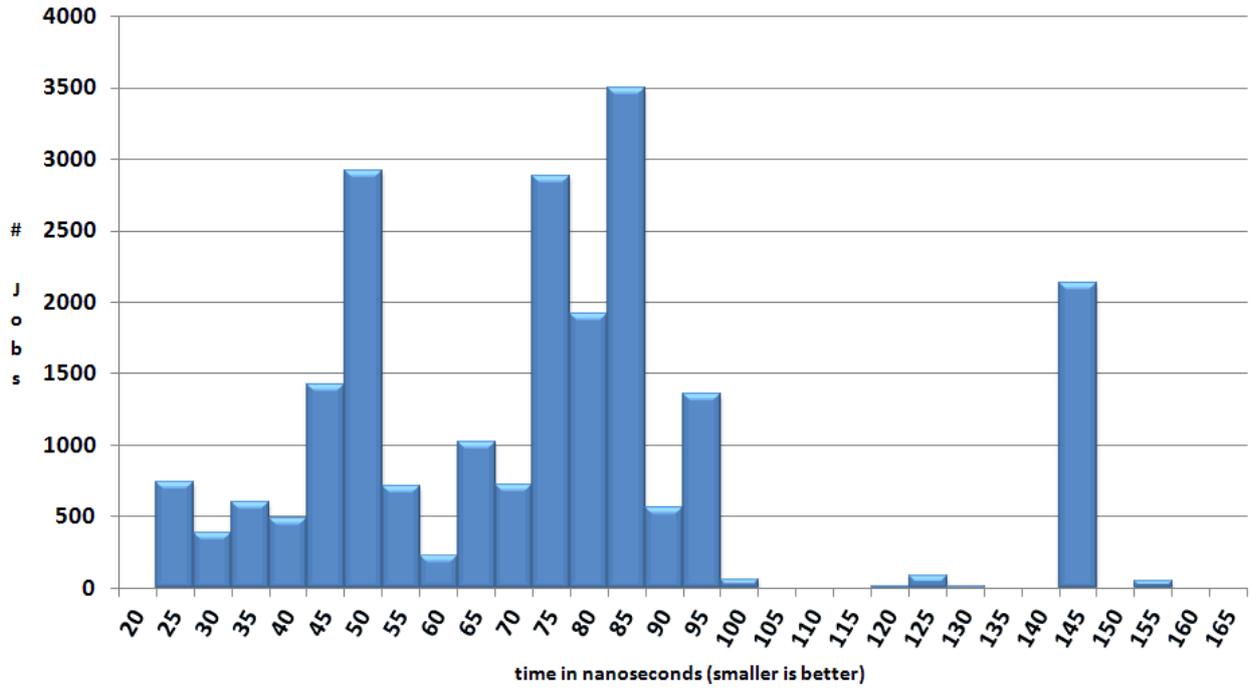
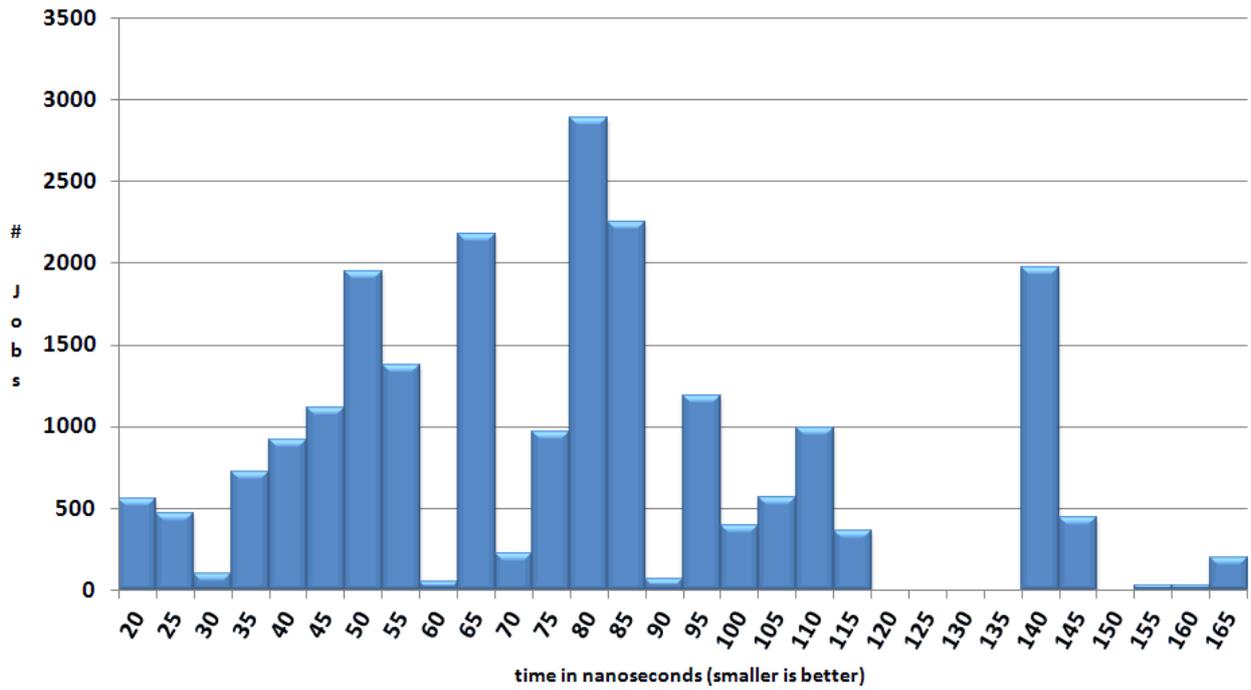

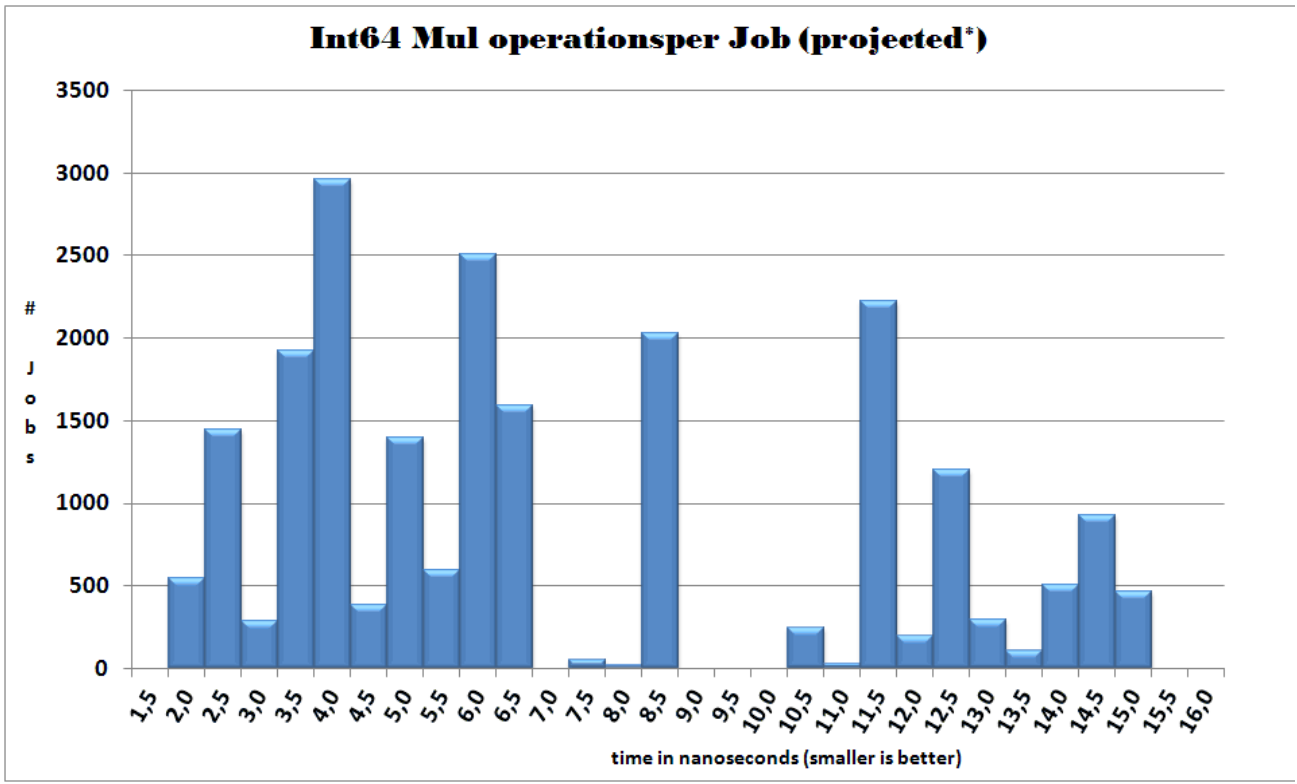

We also took measurements for float and double number operations.

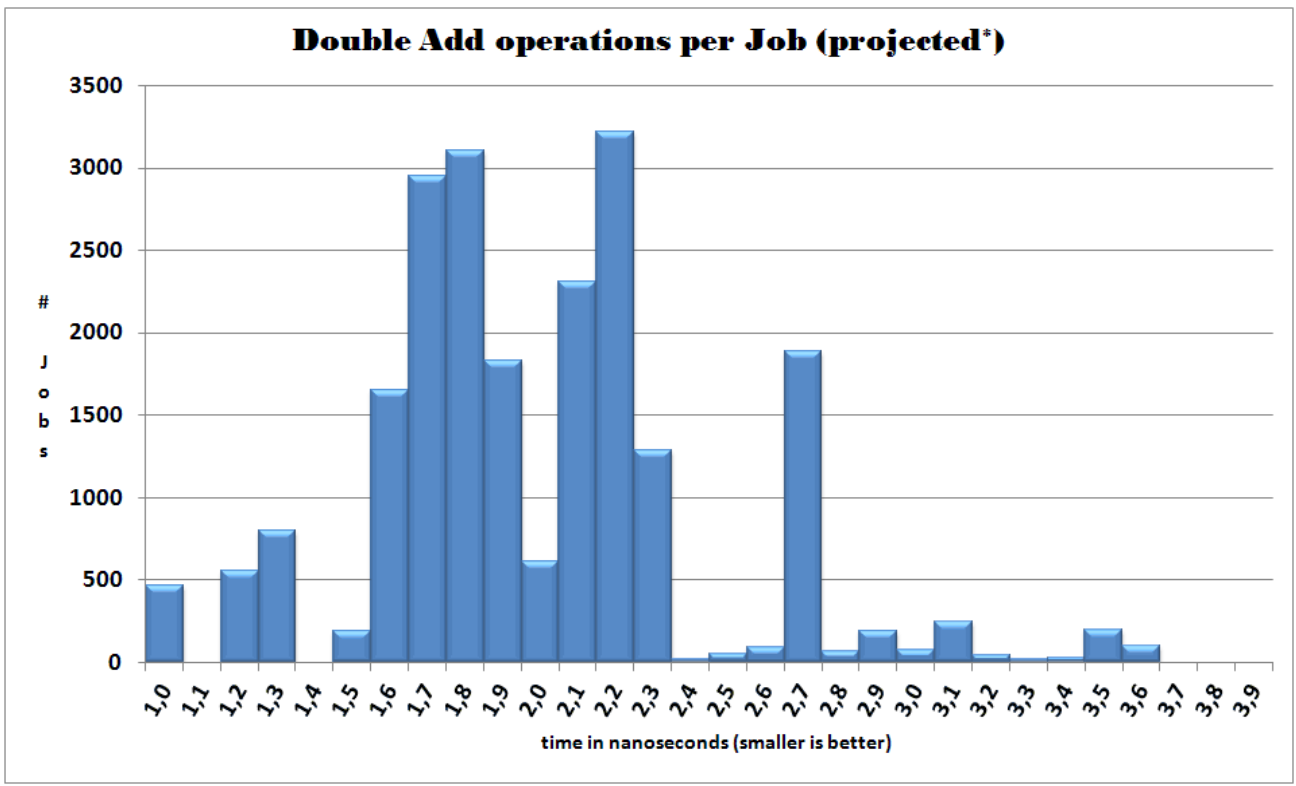

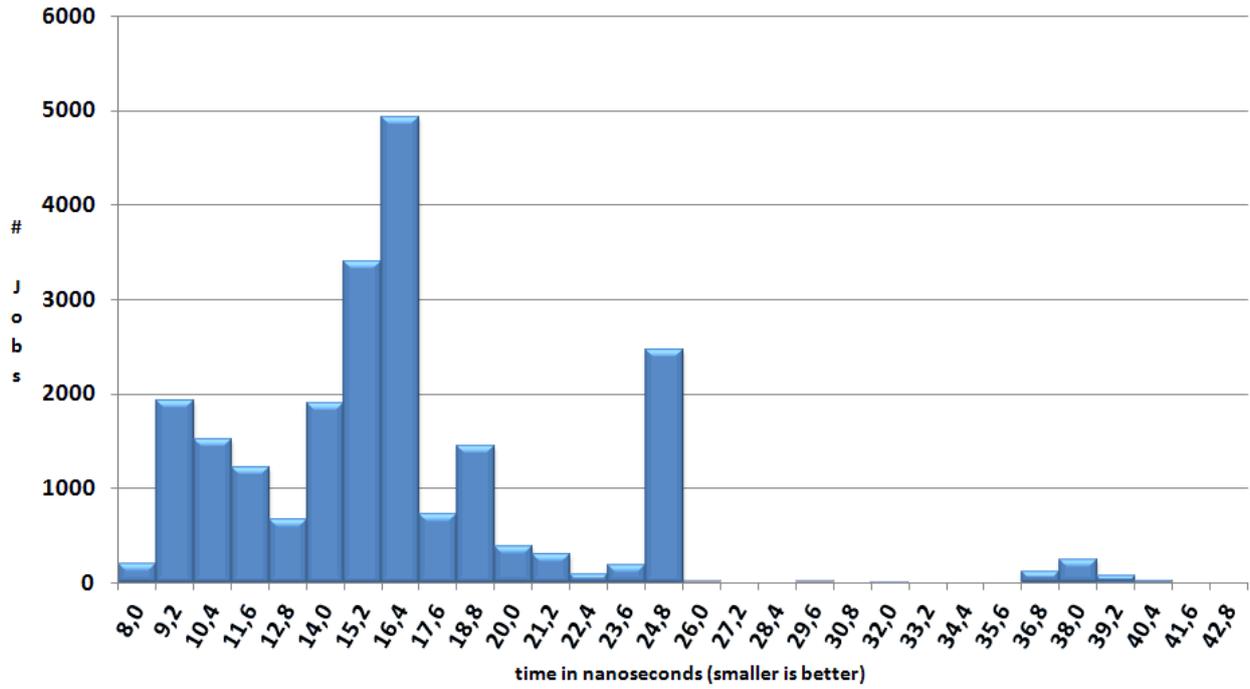
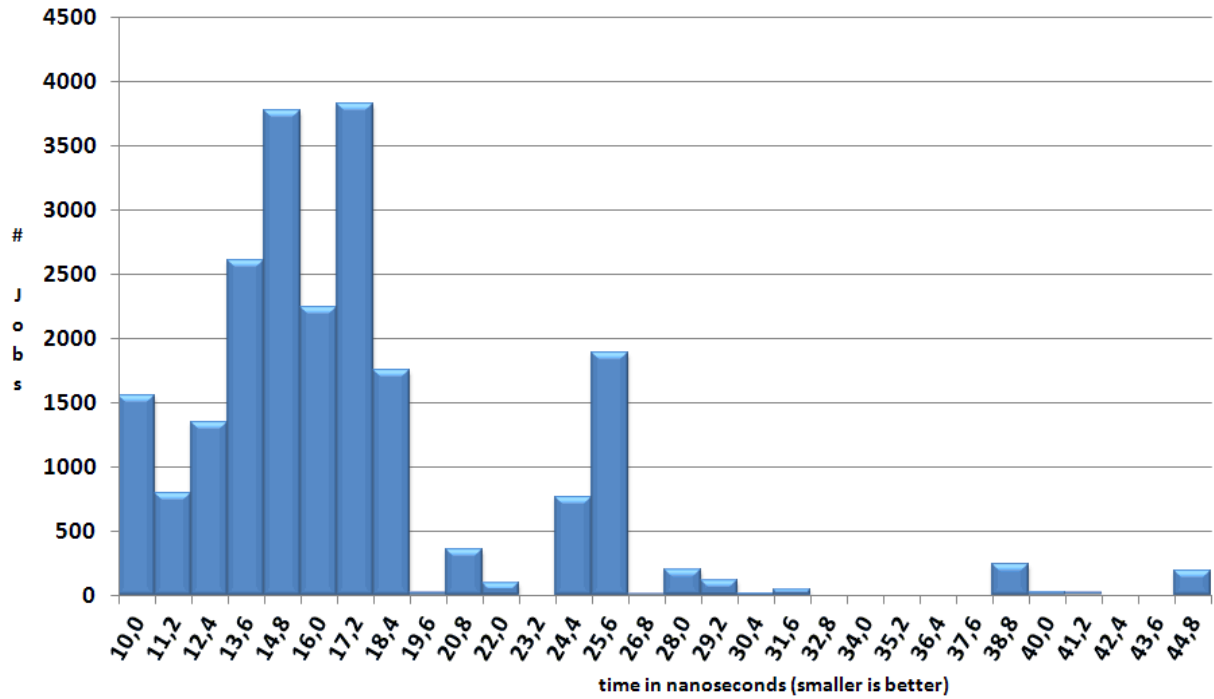

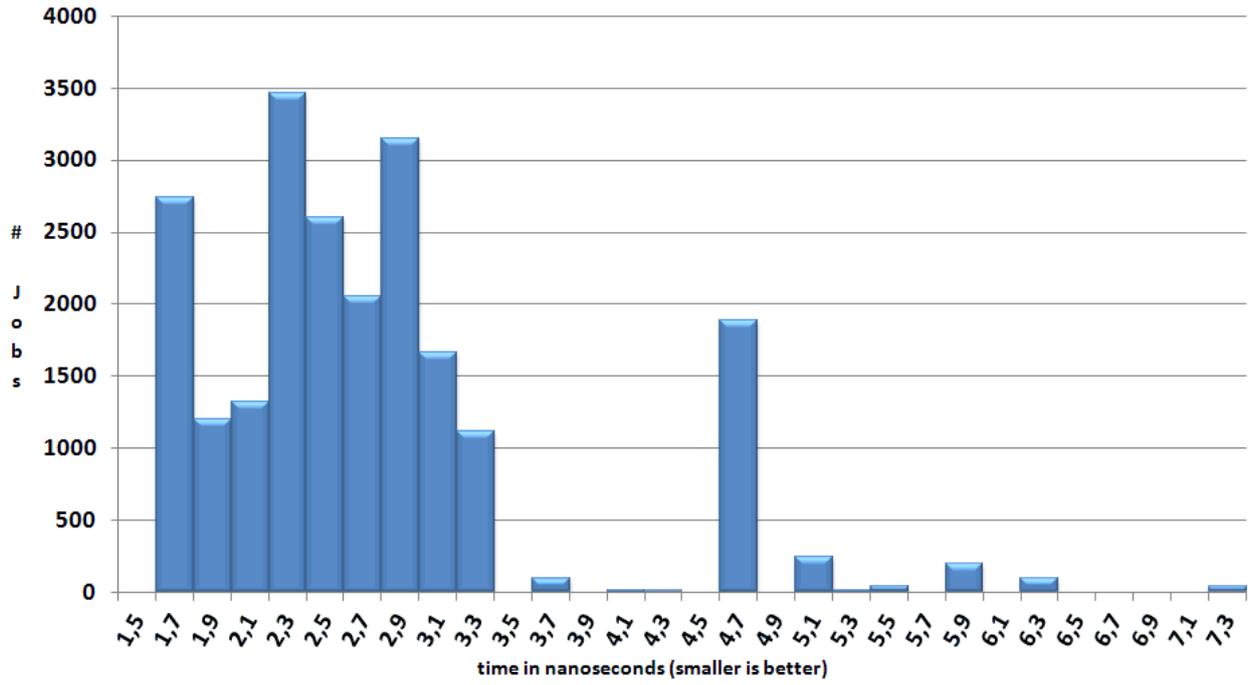
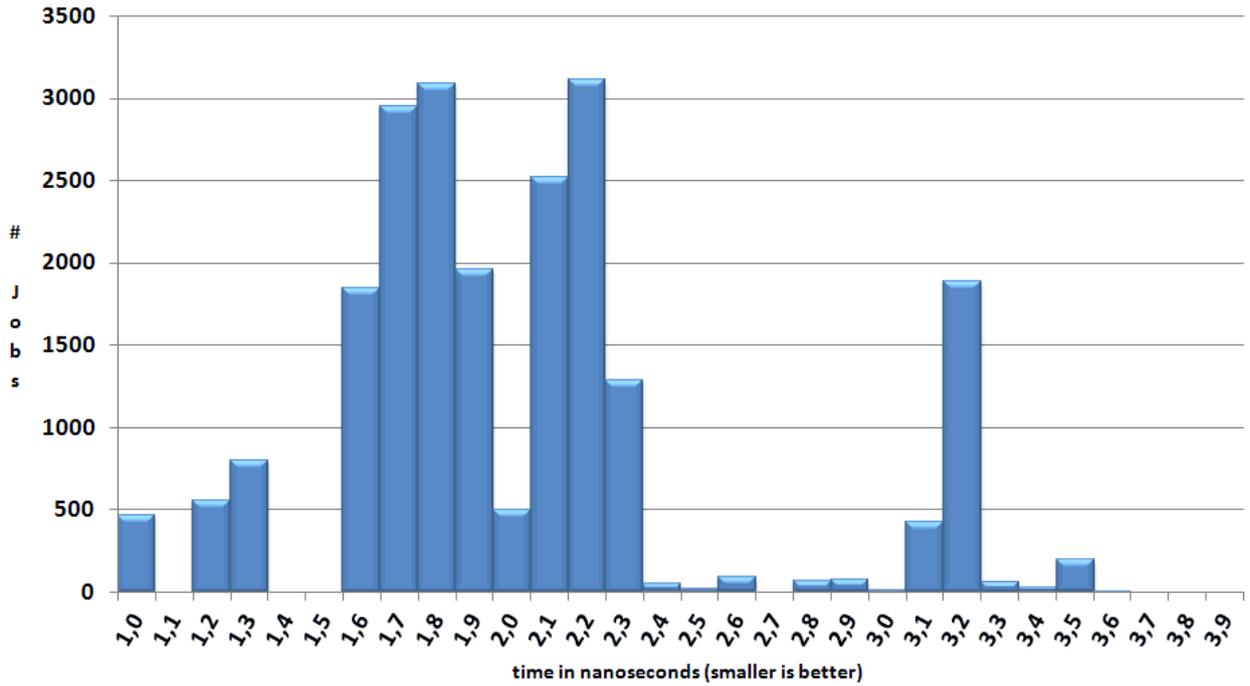

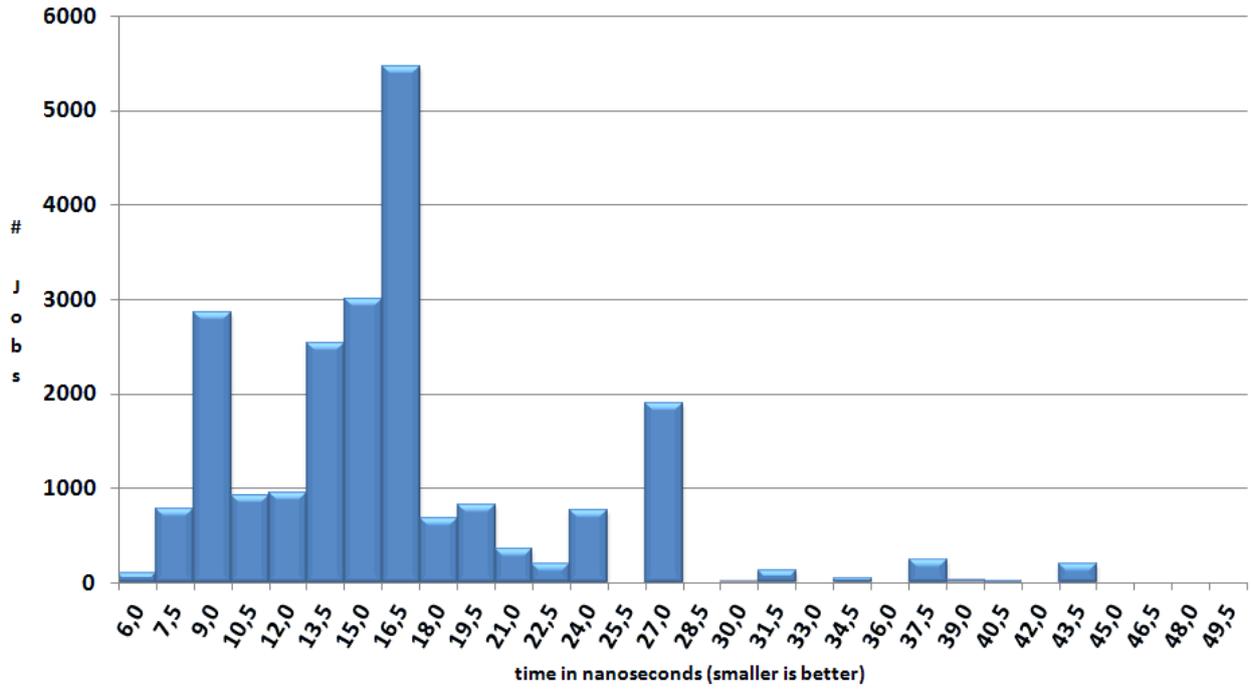

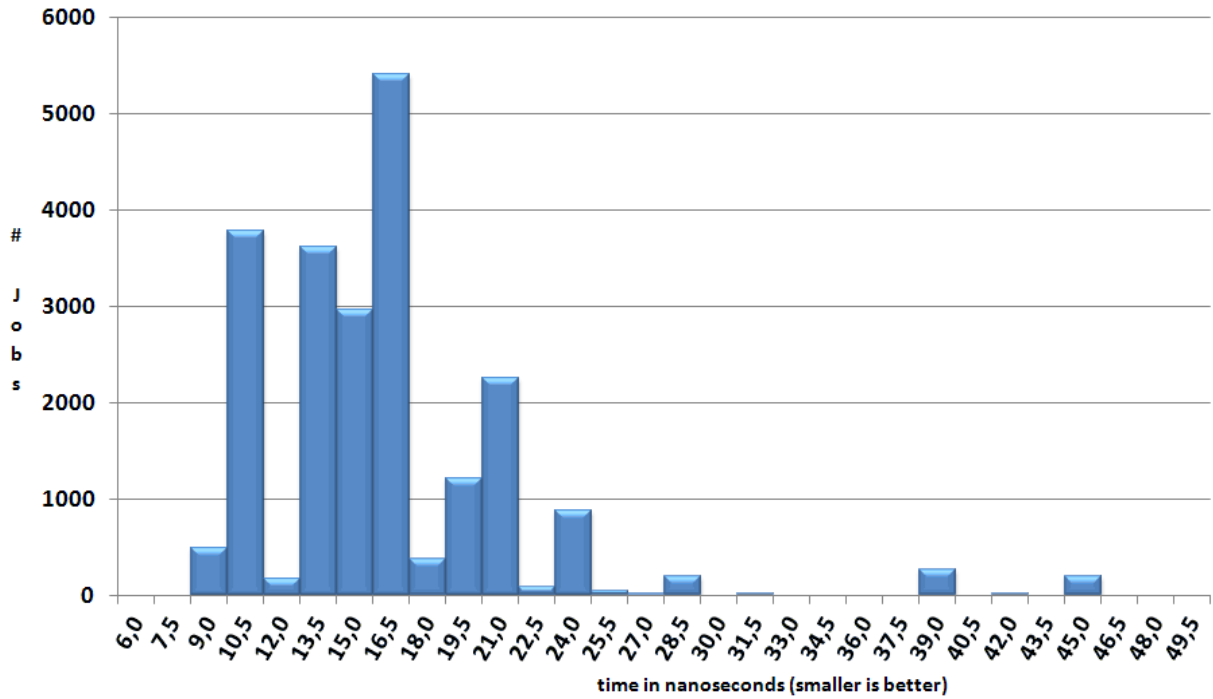

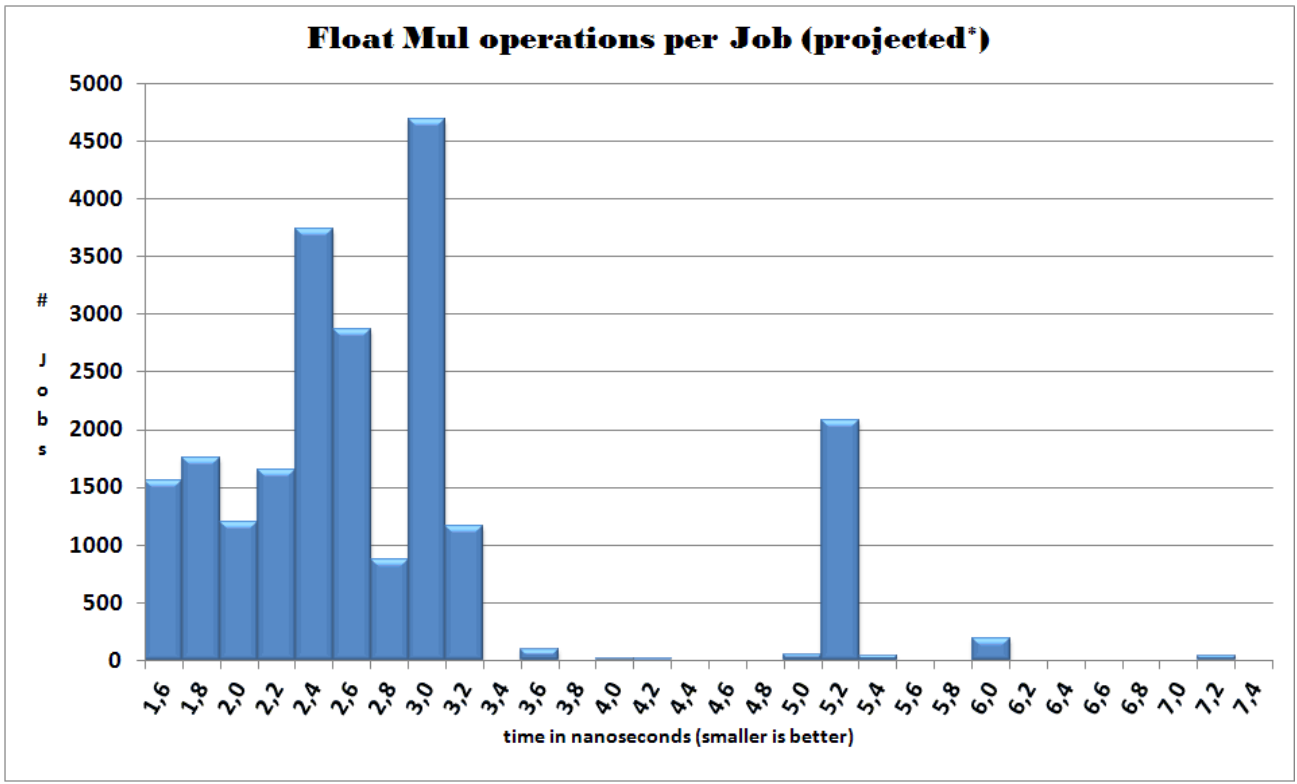

In the next table we see measures about memory latency, which includes L1 and L2 memory. The results are reported in nanoseconds.

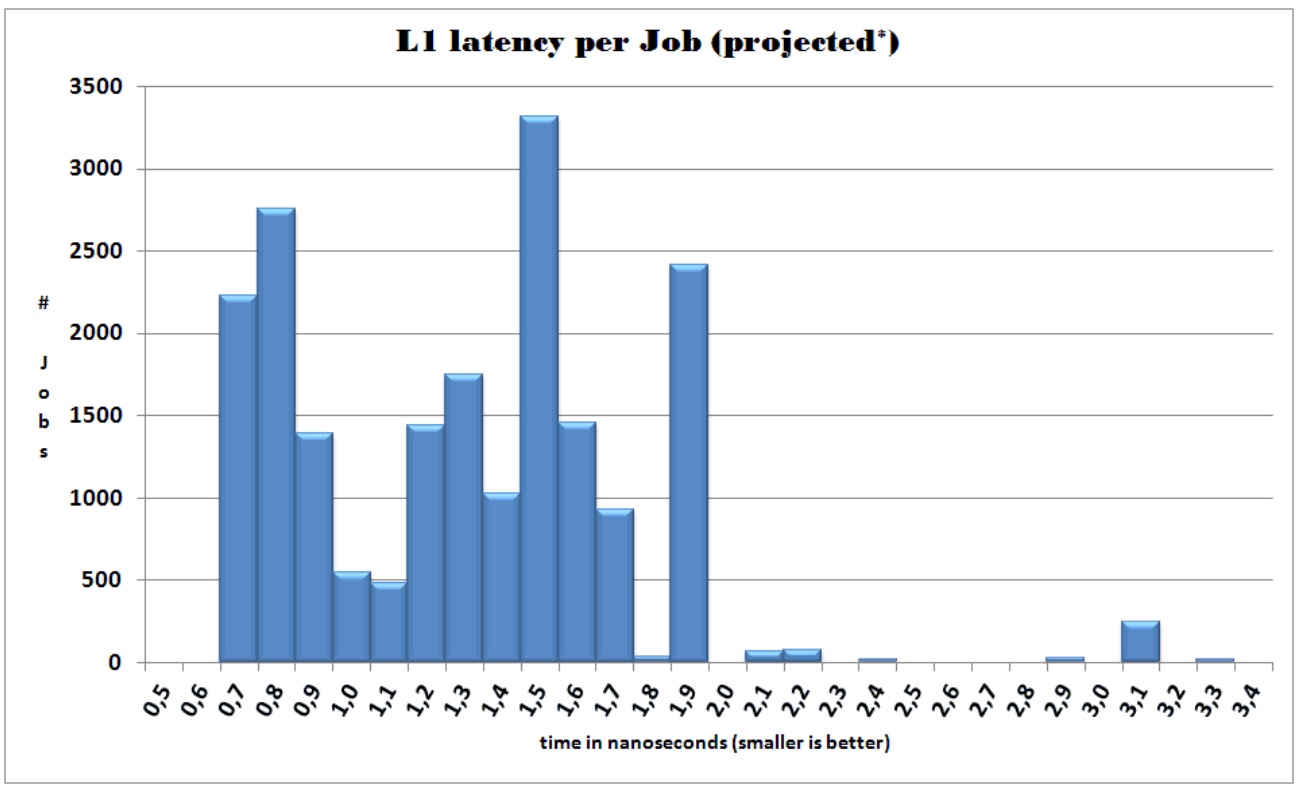

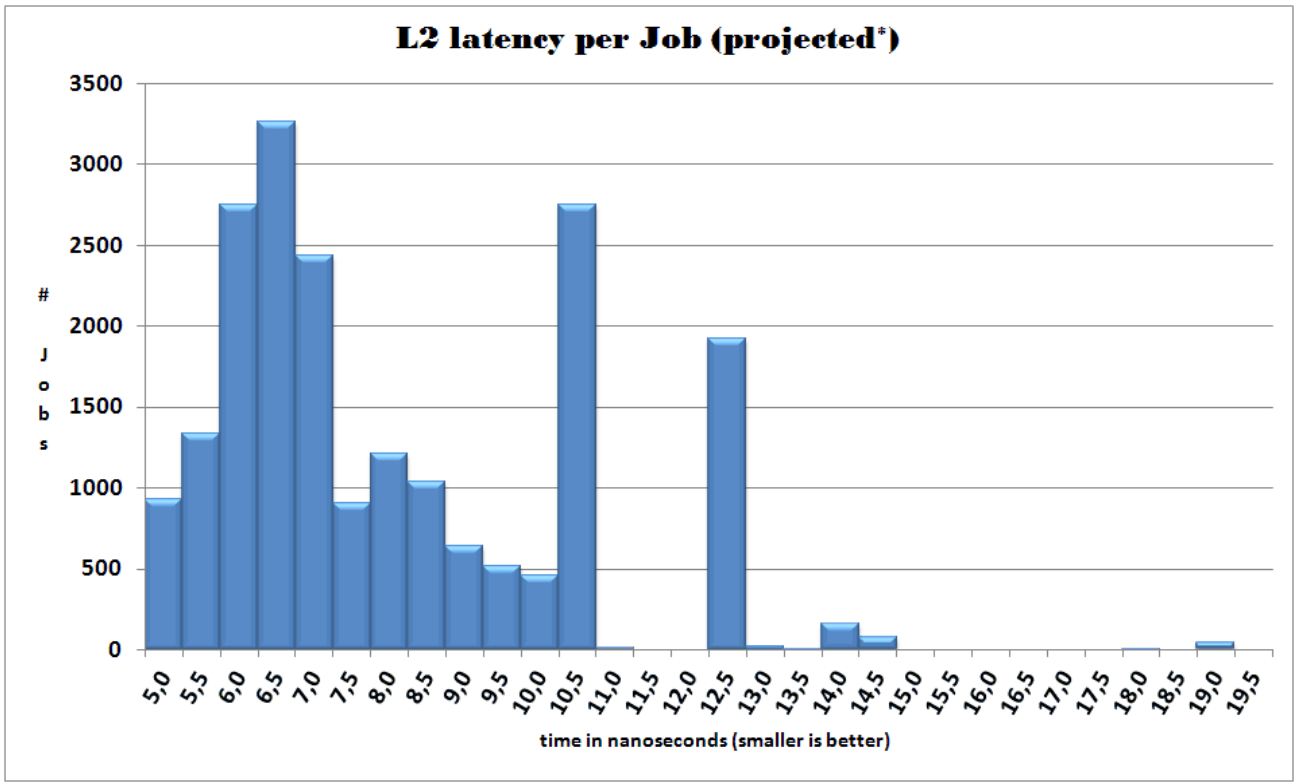

From the LmBench suite we got some other intreasting measurements called stream. Those measurements are also for bandwidth and latency

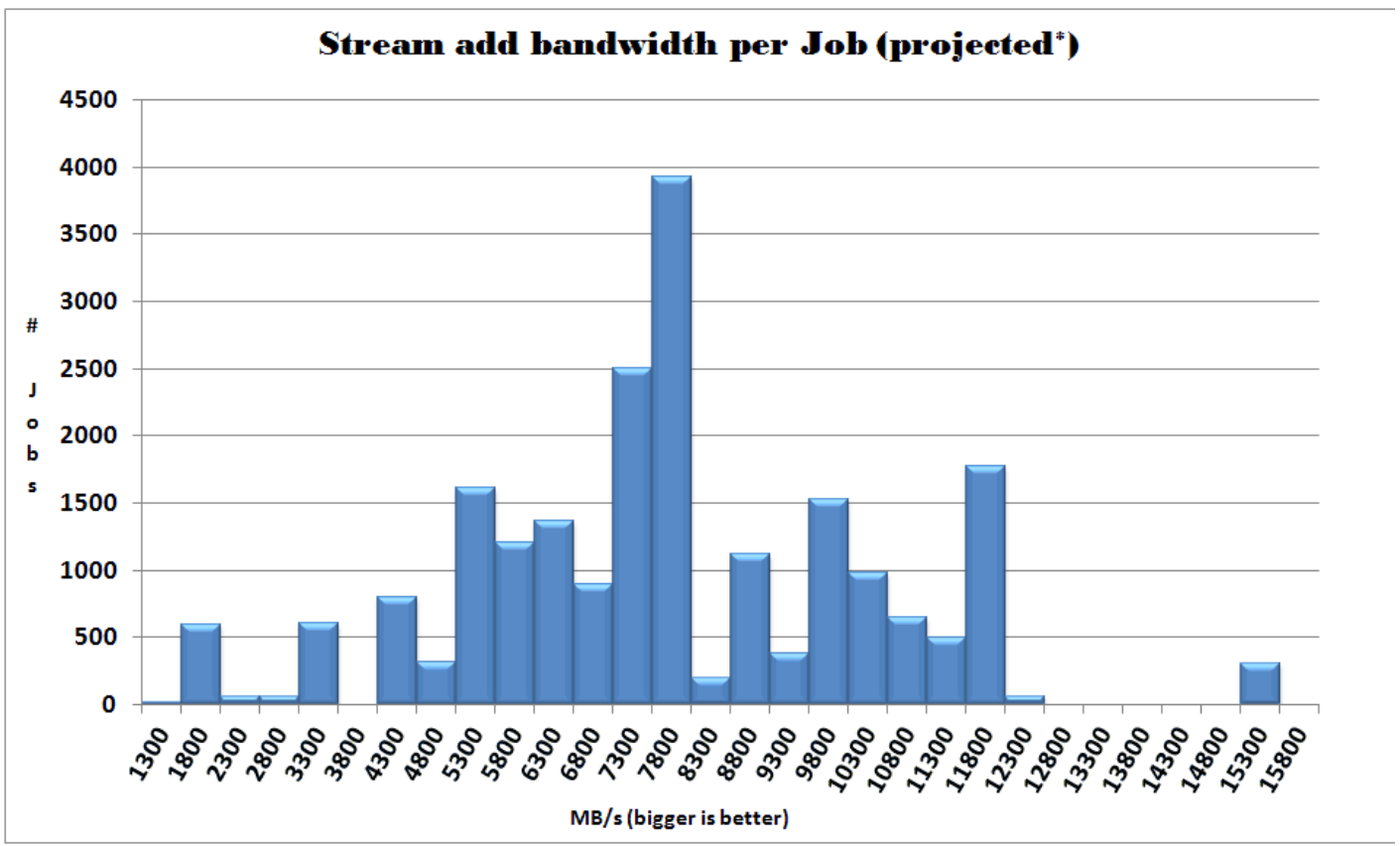

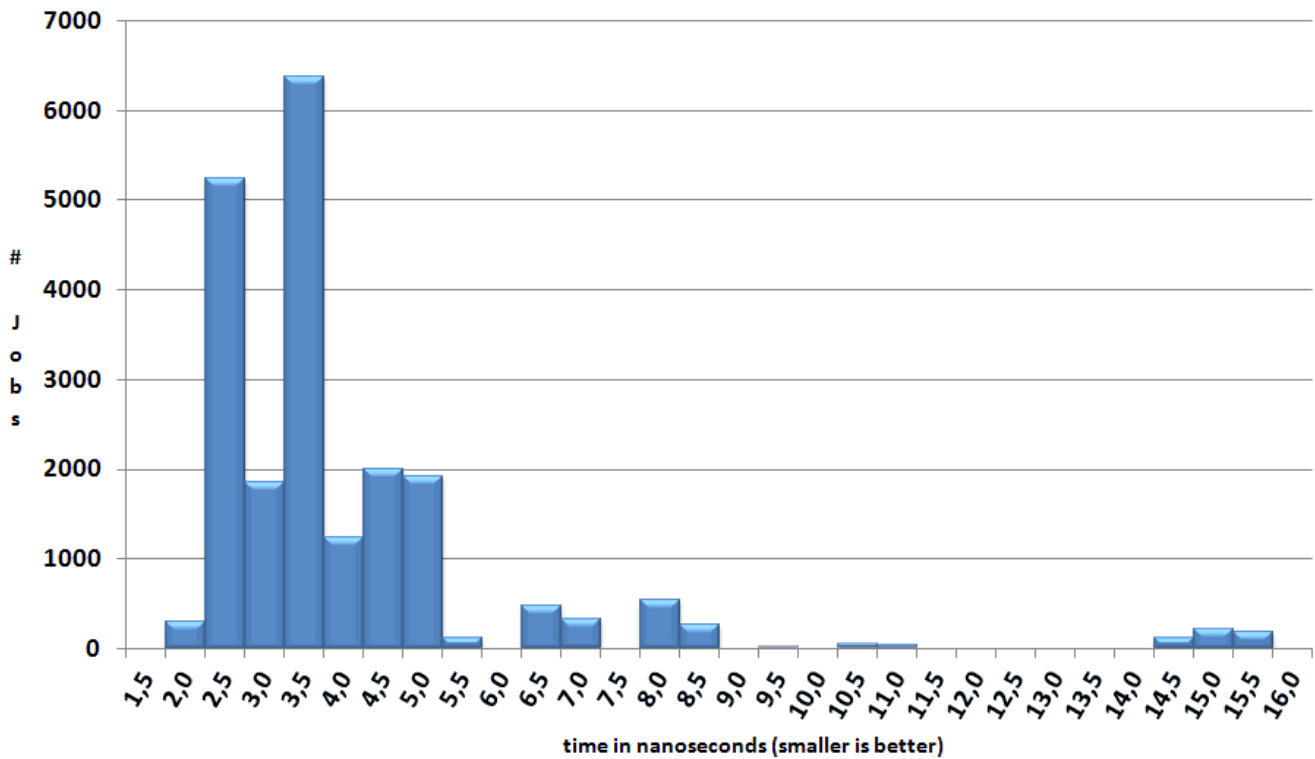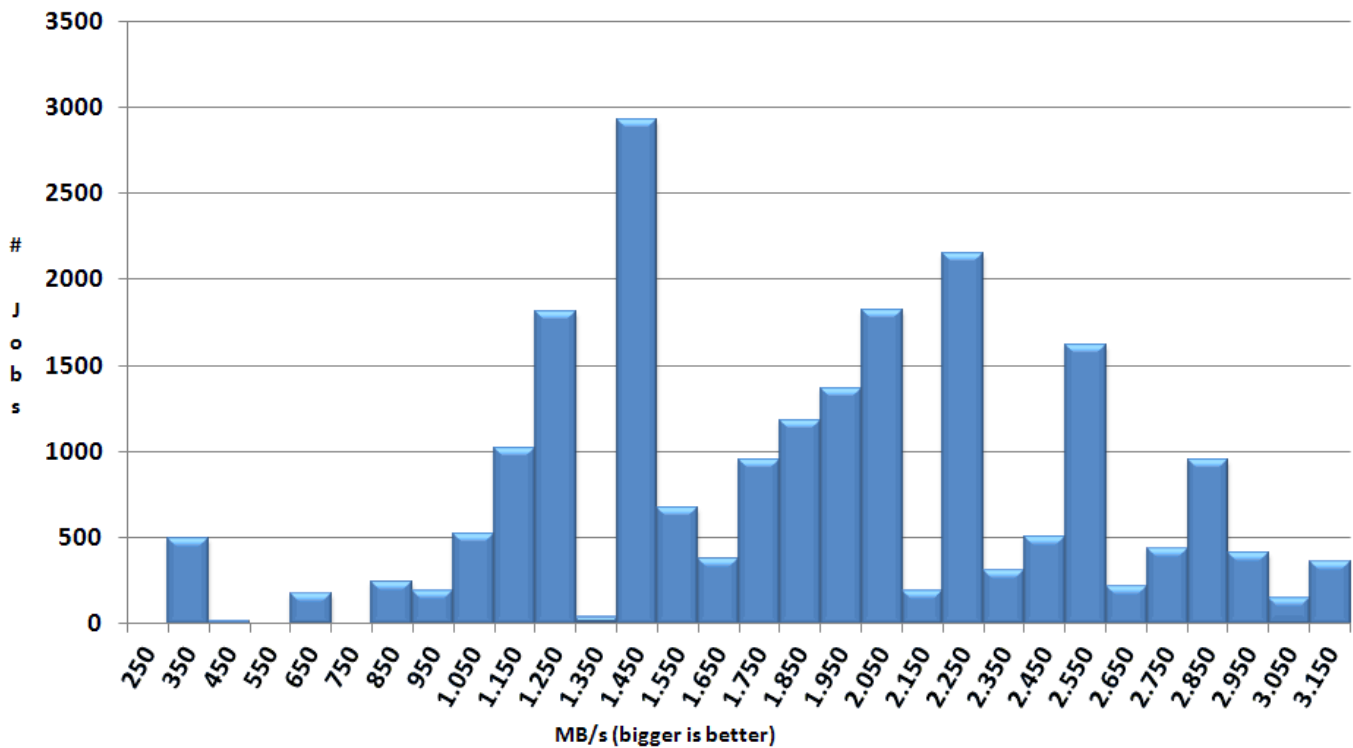

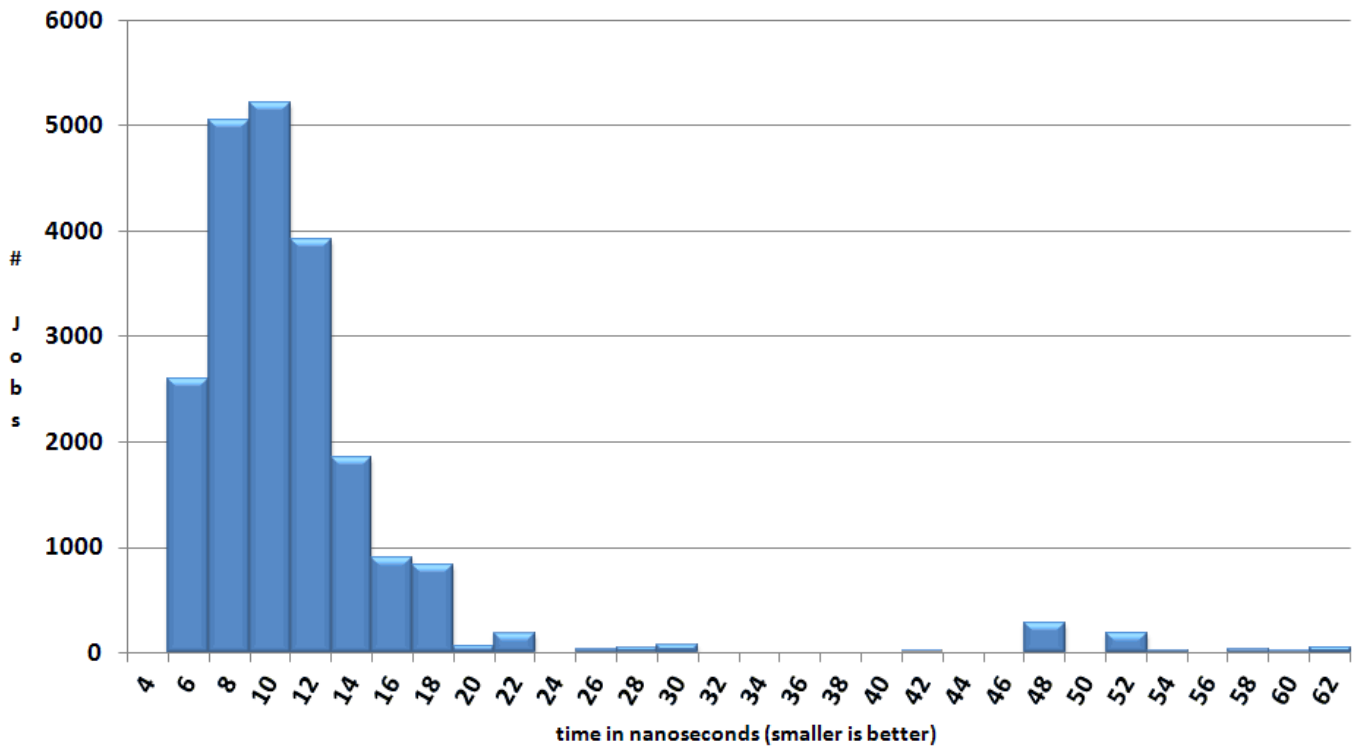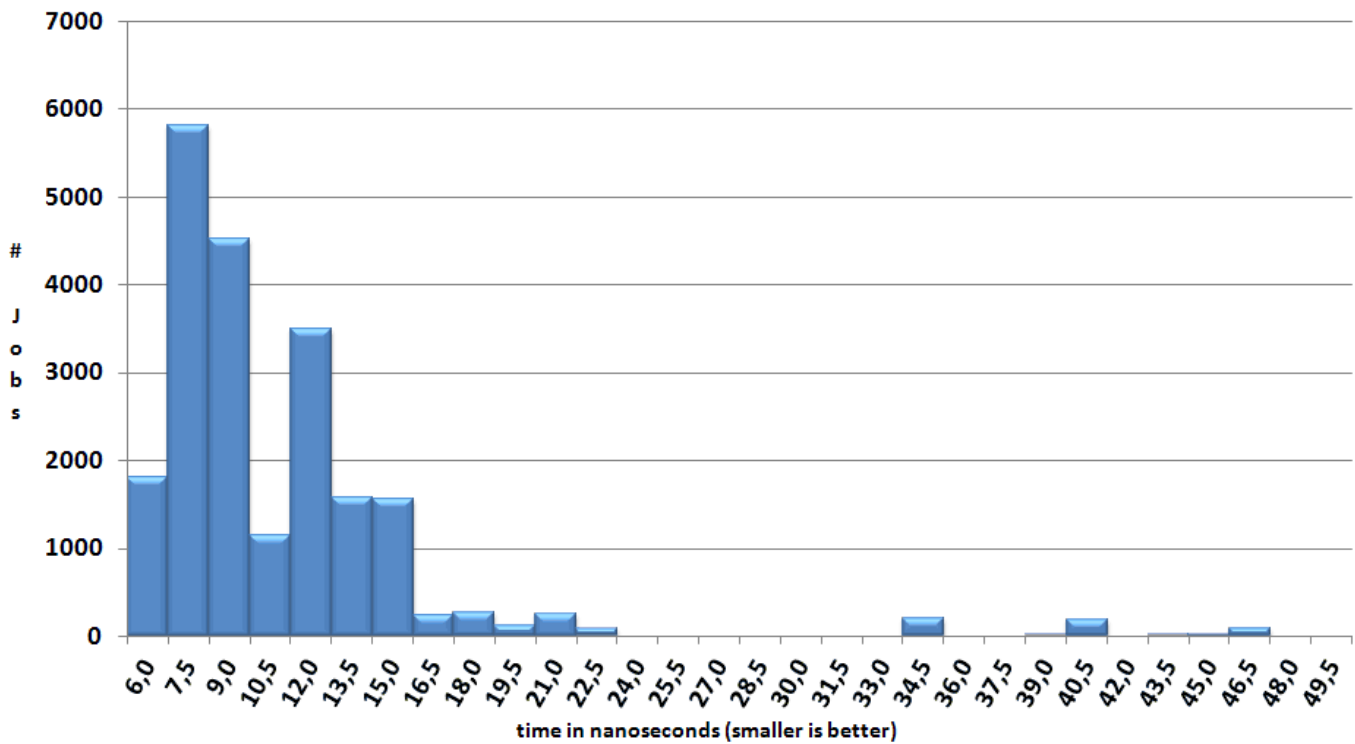

This is the previous chart zoomed in to see the first part that is intrested.

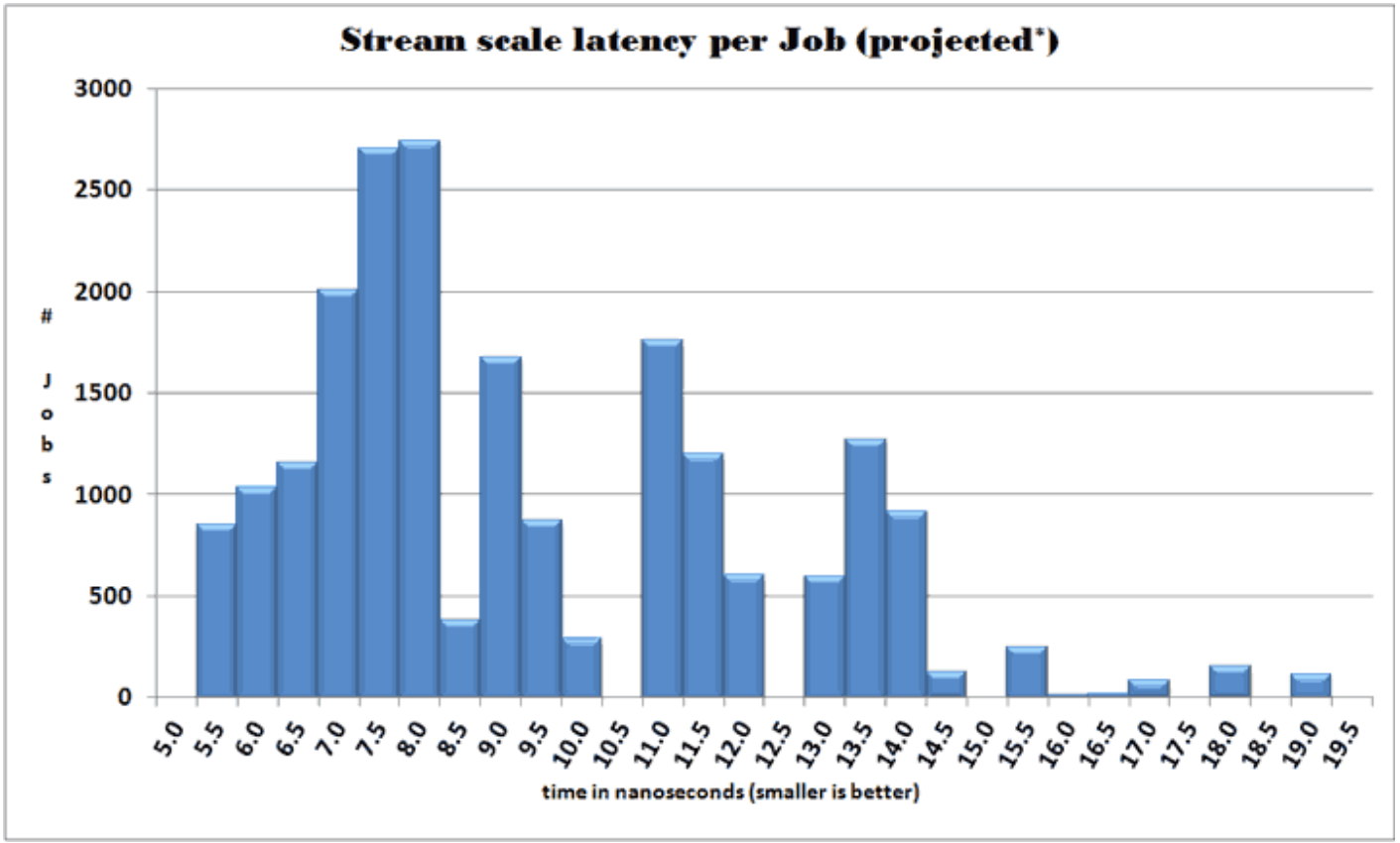

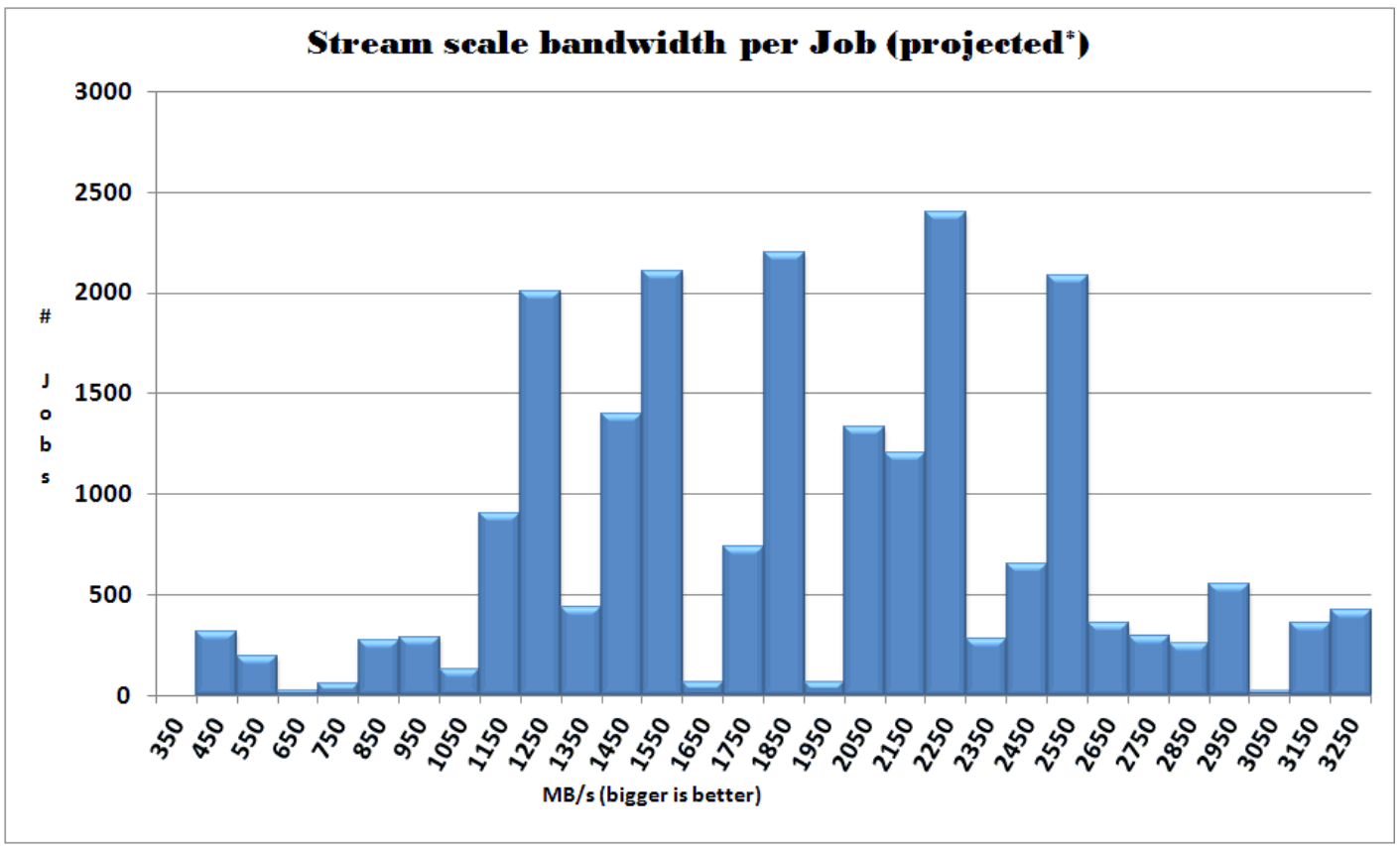

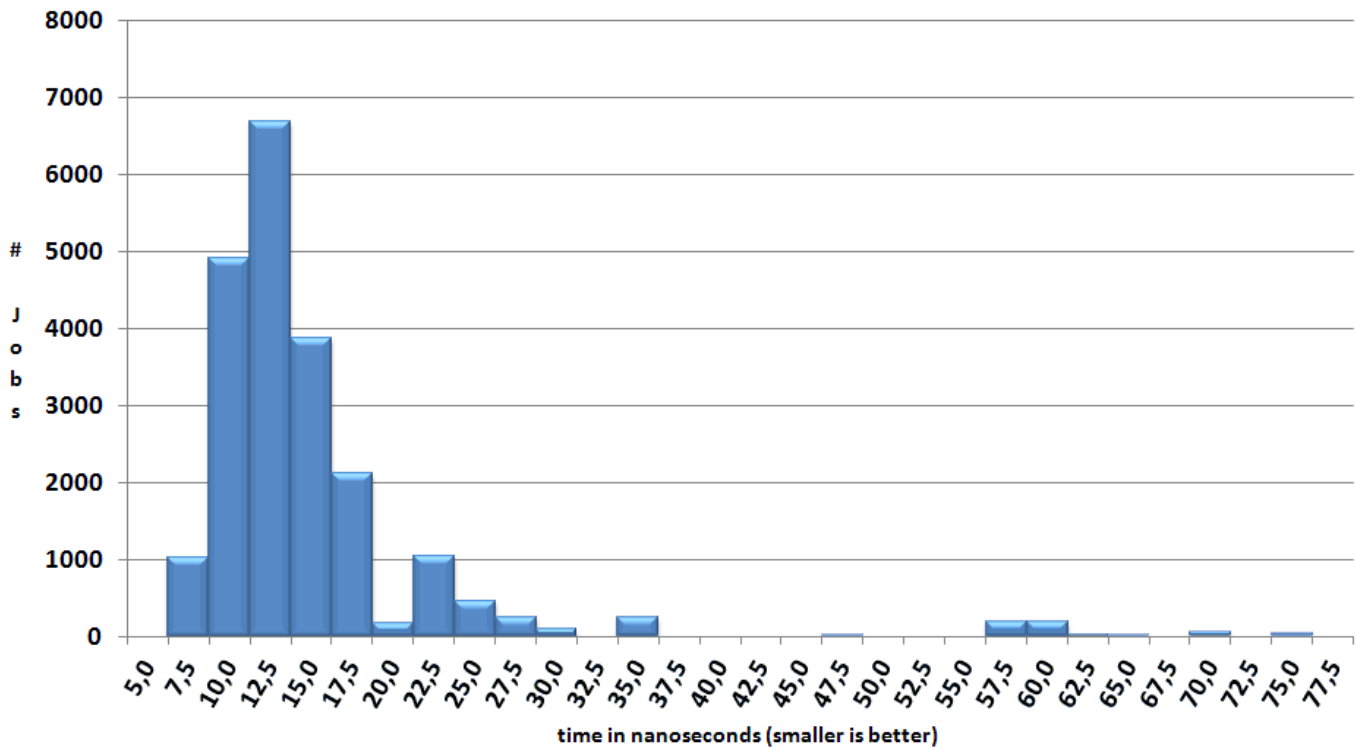

Below is the previous chart zoomed in to see the first part that is intrested.

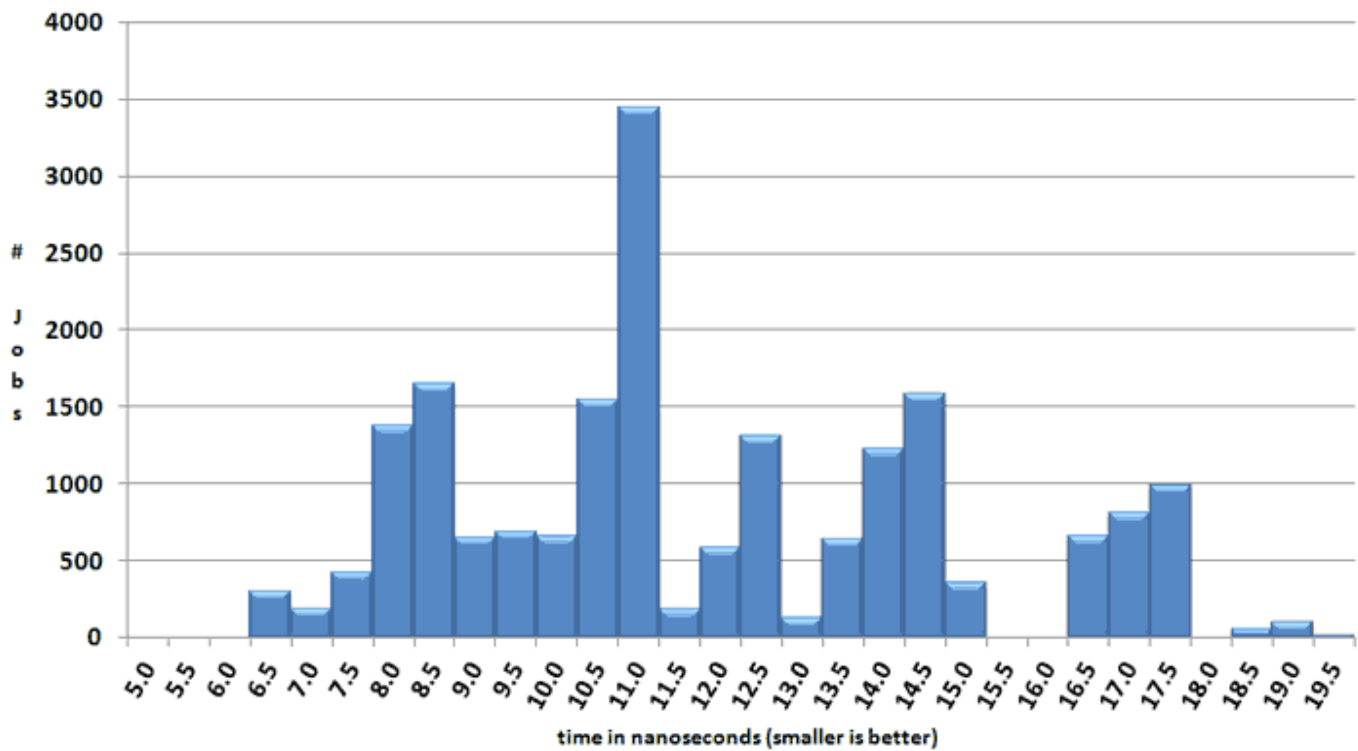

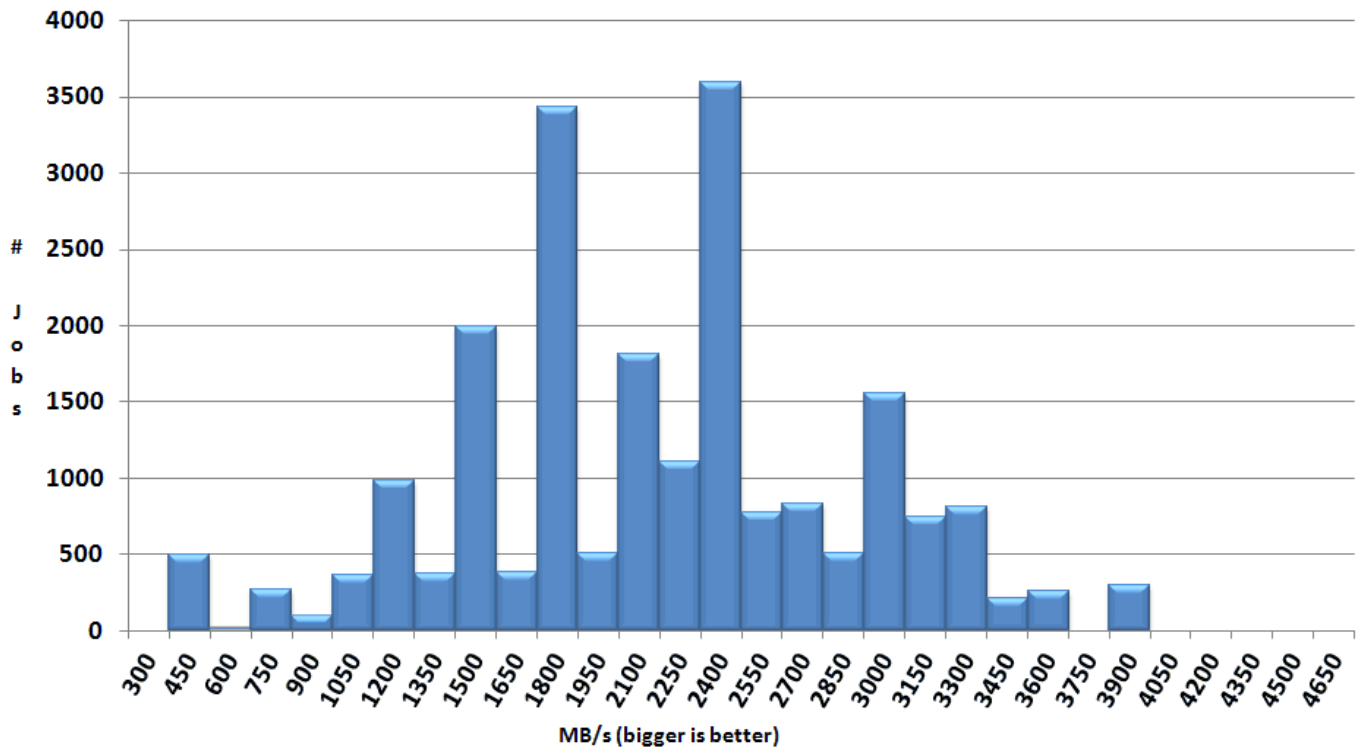

Finally we got also stream2 measurements from lmbench.

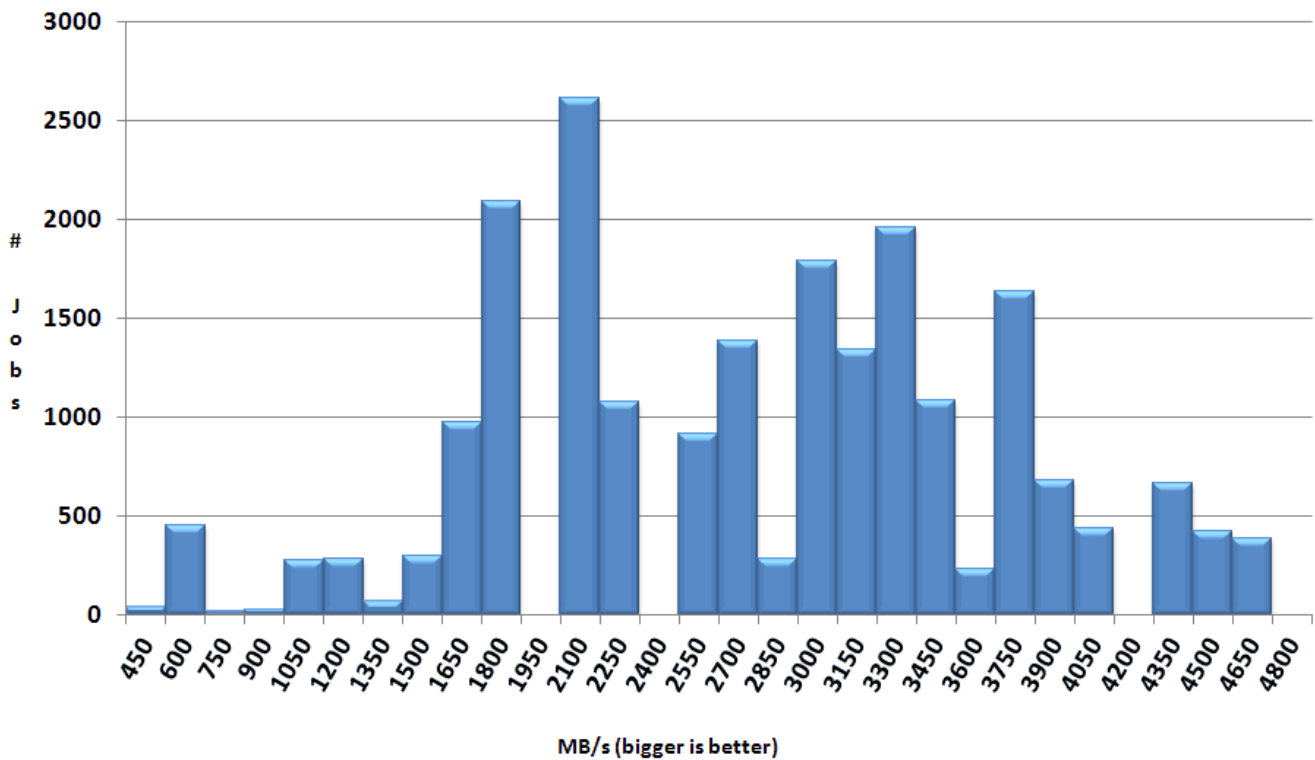

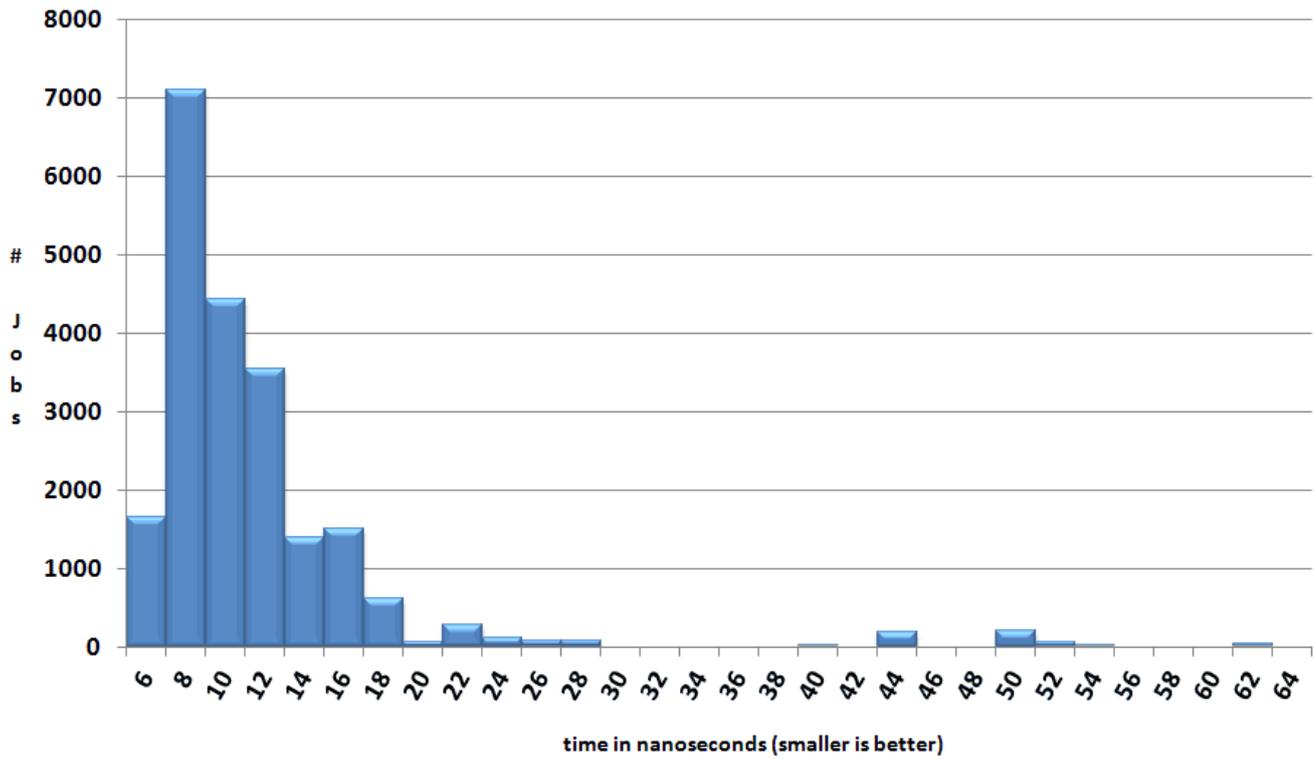

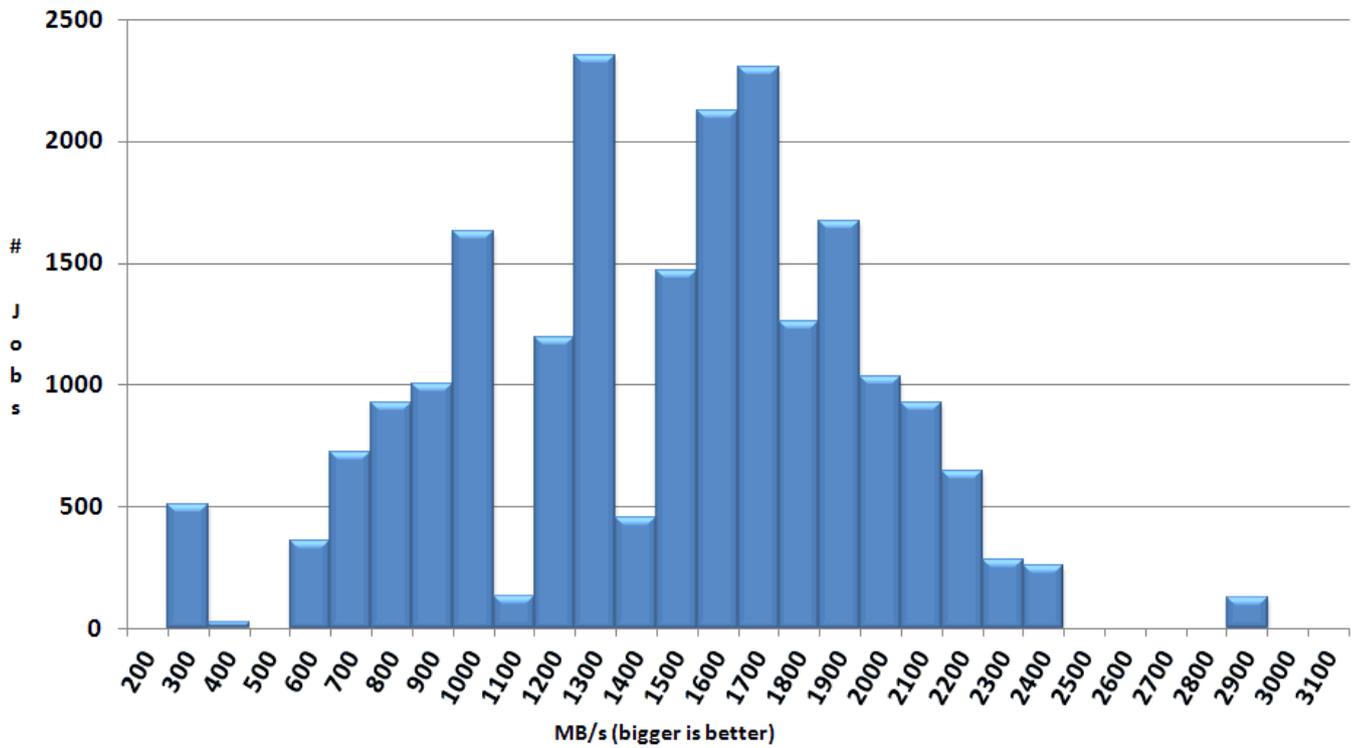

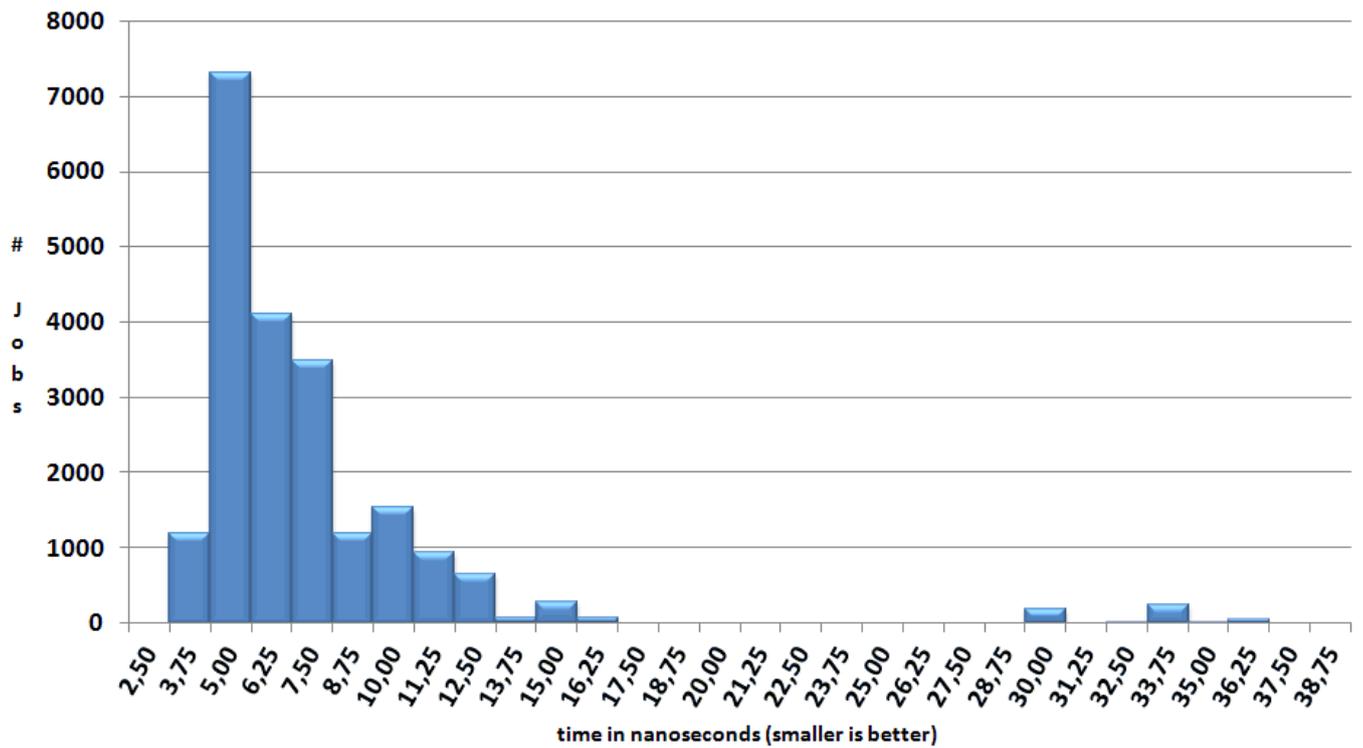
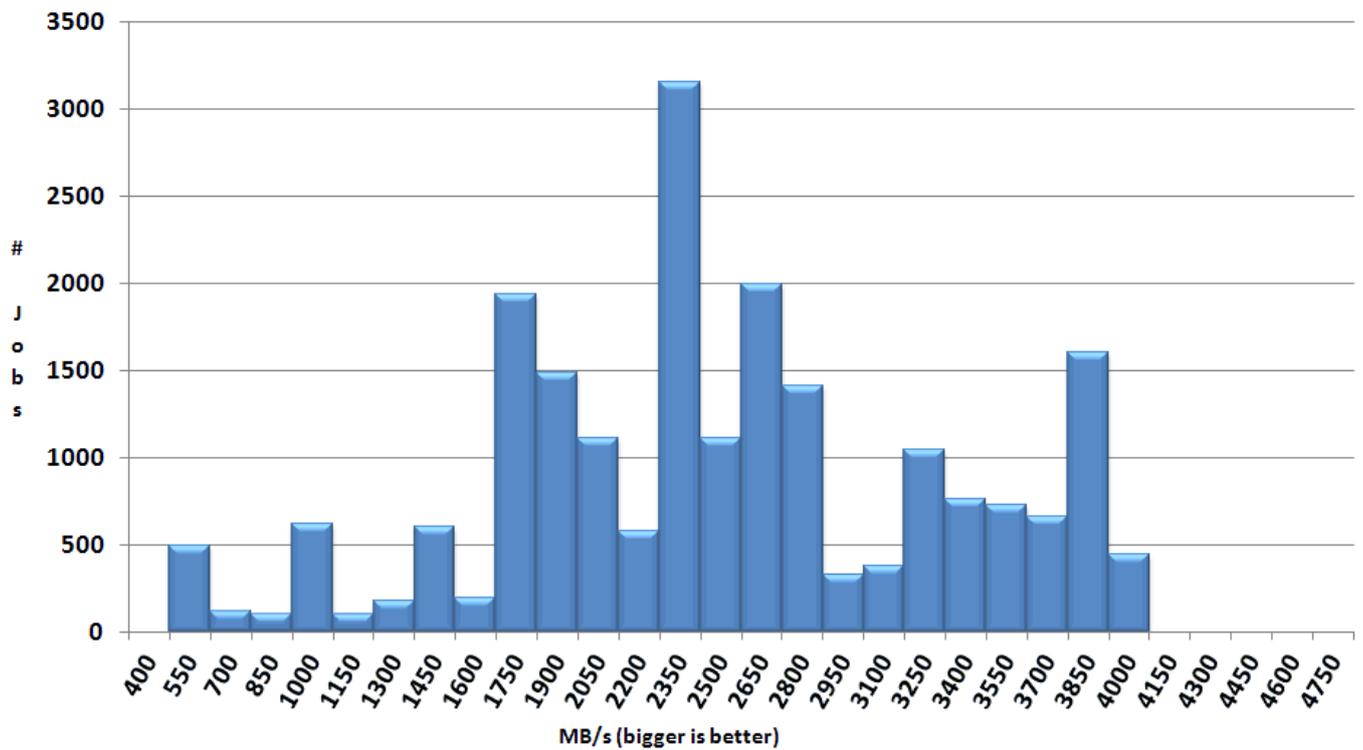

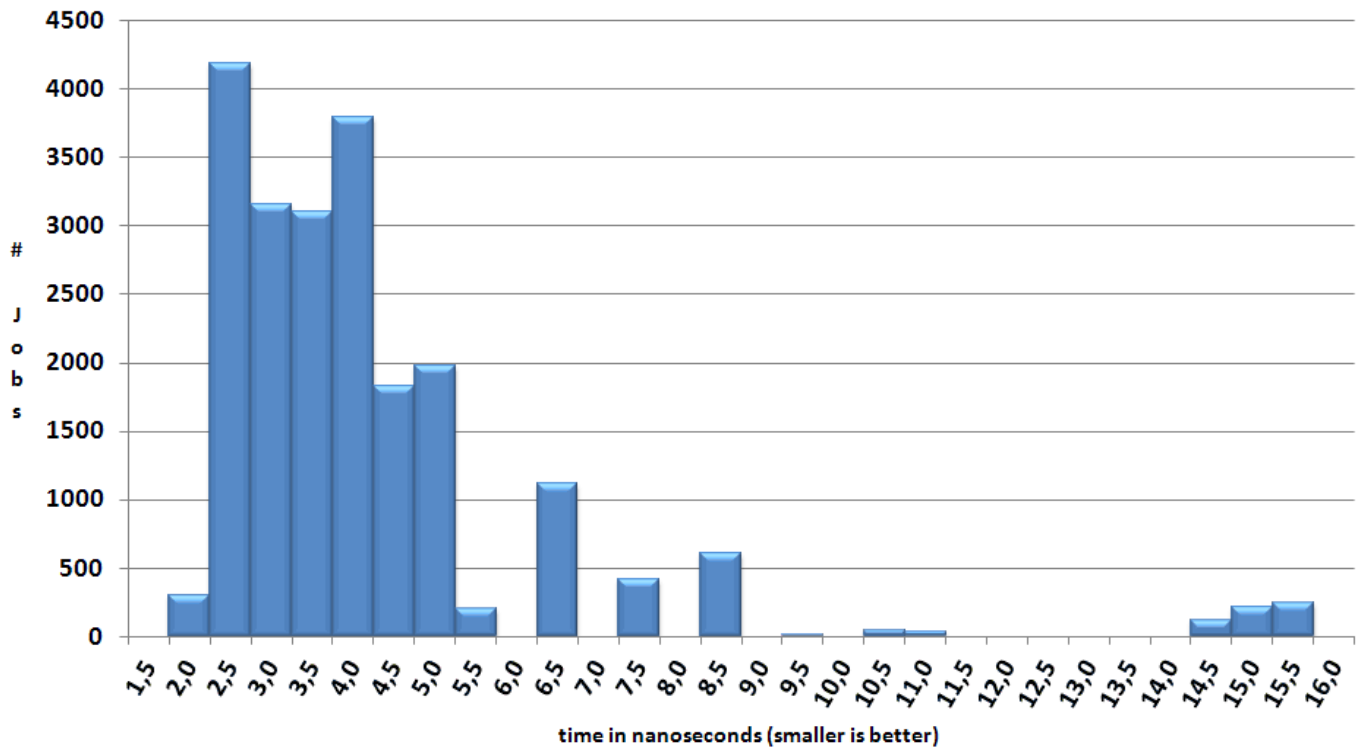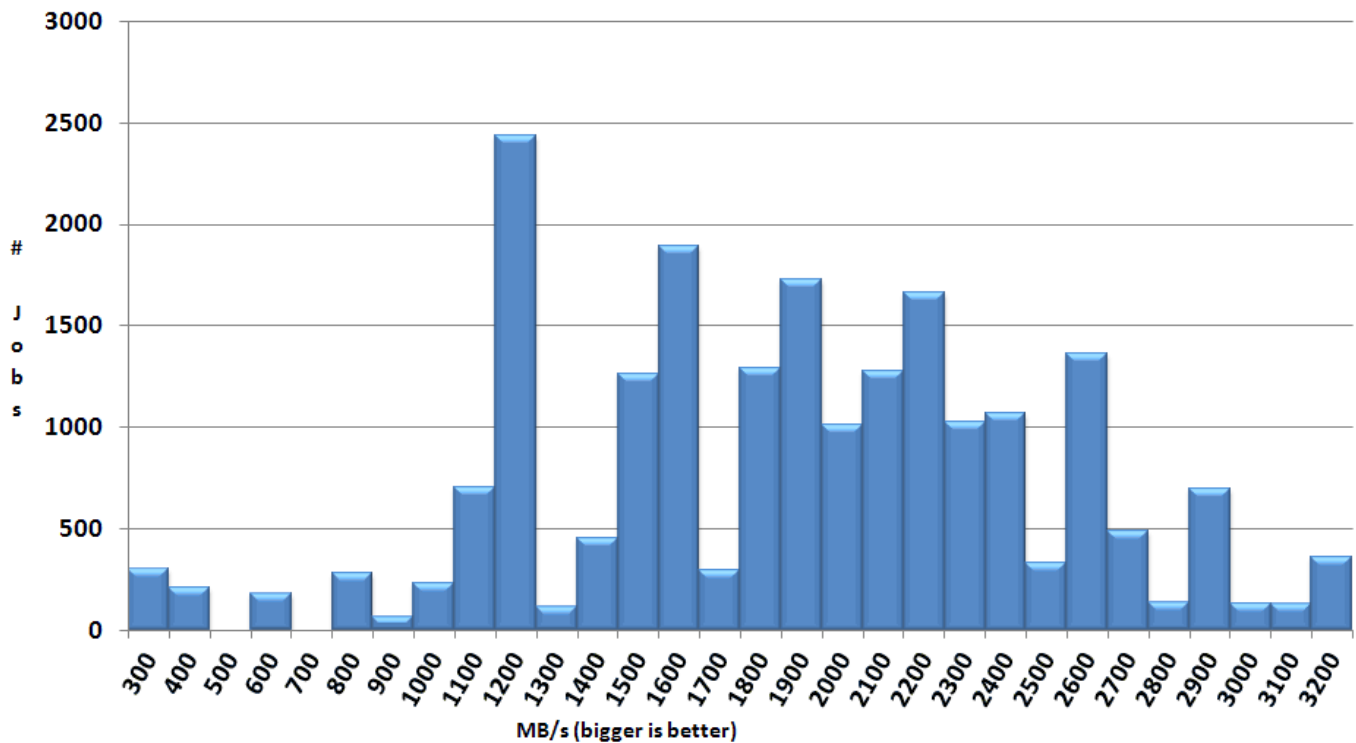

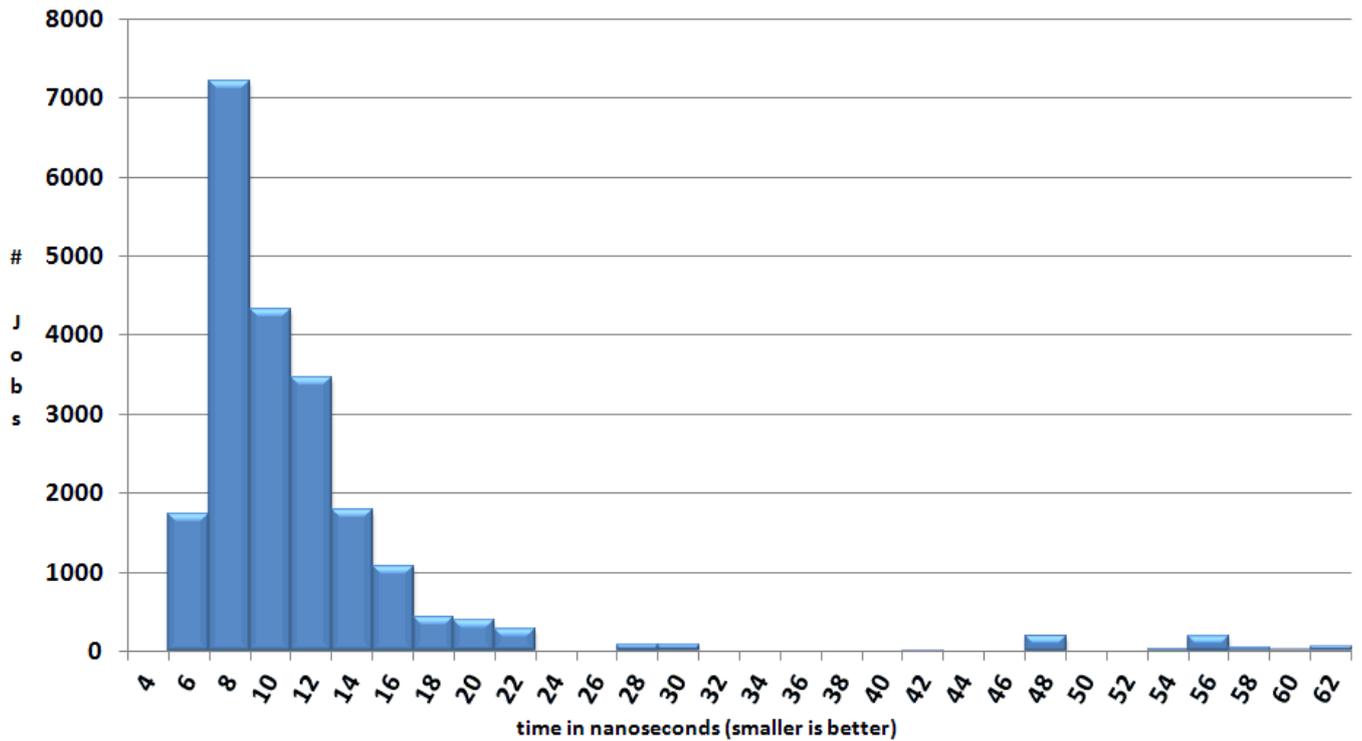

Below is the previous chart zoomed in to see the first part that is intrested.

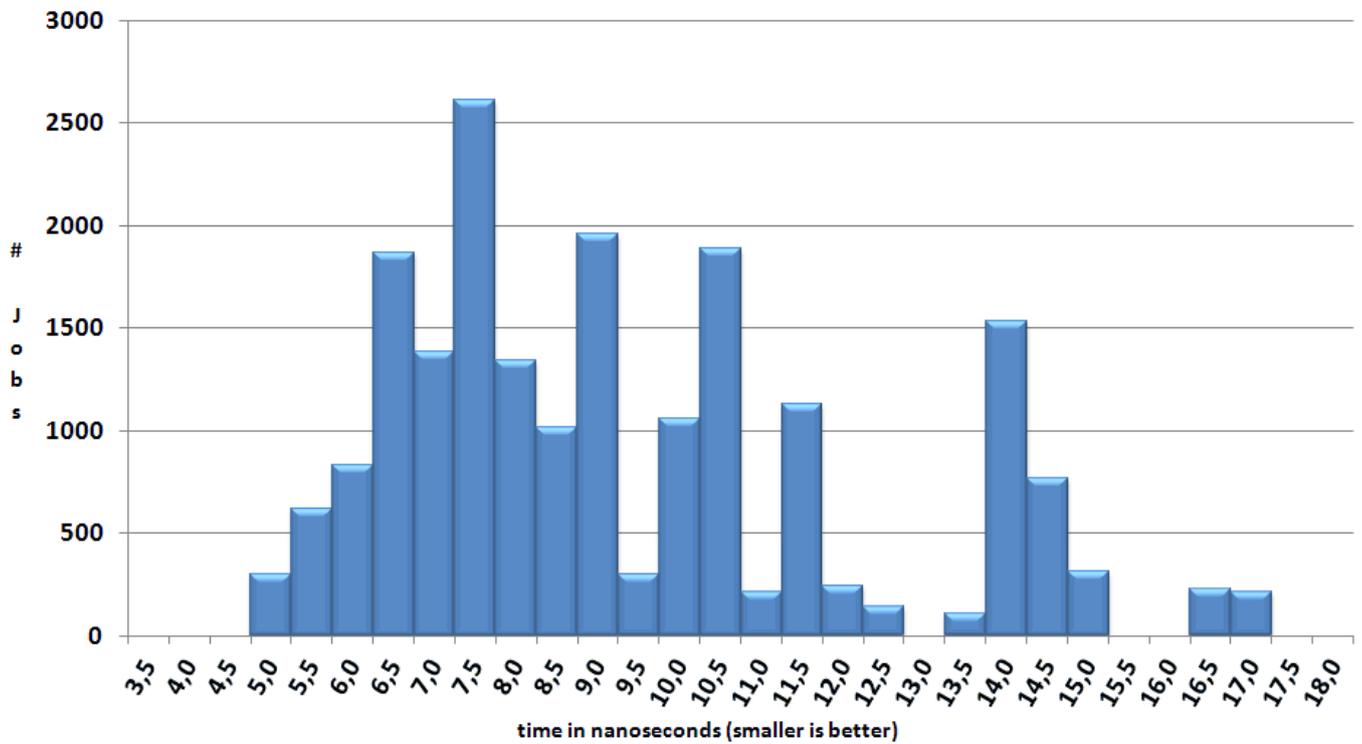